\documentclass[preprint,trackchanges]{aastex631}
\received{November 5, 2022}
\accepted{December 16, 2022}

\shortauthors{Kotani et al.}
\graphicspath{{./}{figures/}}

\begin{document}

\title{Unified Relationship between Cold Plasma Ejections and Flare Energies Ranging from Solar Microflares to Giant Stellar Flares}

\author[0000-0002-5112-2740]{Yuji Kotani}
\affiliation{Astronomical Observatory, Kyoto University, Sakyo, Kyoto 606-8502, Japan}
\correspondingauthor{Yuji Kotani}
\email{kotani@kusastro.kyoto-u.ac.jp}
%
%
\author{Kazunari Shibata}
\affiliation{Astronomical Observatory, Kyoto University, Sakyo, Kyoto 606-8502, Japan}
\affiliation{Department of Environmental Systems Science, Faculty of Science and Engineering,\\
Doshisha University,
1-3, Tatara Miyakodani, Kyotanabe City, Kyoto, 610-0394, Japan}
\author{Takako T.  Ishii}
\affiliation{Astronomical Observatory, Kyoto University, Sakyo, Kyoto 606-8502, Japan}

\author[0000-0003-1072-3942]{Daiki Yamasaki}
\affiliation{Astronomical Observatory, Kyoto University, Sakyo, Kyoto 606-8502, Japan}

\author{Kenichi Otsuji}
\affiliation{Space Environment Laboratory, Applied Electromagnetic Research Institute, \\
National Institute of Information and Communications Technology, Koganei, Tokyo 184-8795, Japan.
}

\author[0000-0001-8689-3564]{Kiyoshi Ichimoto}
\affiliation{Astronomical Observatory, Kyoto University, Sakyo, Kyoto 606-8502, Japan}

\author{Ayumi Asai}
\affiliation{Astronomical Observatory, Kyoto University, Sakyo, Kyoto 606-8502, Japan}

%
%
%



\begin{abstract}
We often find spectral signatures of chromospheric cold plasma ejections accompanied by flares in a wide range of spatial scales in the solar and stellar atmospheres.
However, the relationship between physical quantities (such as mass, kinetic energy, and velocity) of cold ejecta and flare energy has not been investigated in a unified manner for the entire range of flare energies to date.
This study analyzed the spectra of cold plasma ejections associated with  small-scale flares and solar flares (energy $10^{25}-10^{29}\,\mathrm{erg}$) to supply smaller energy samples.
We performed H$\alpha$ imaging spectroscopy observation by the \textit{Solar Dynamics Doppler Imager} on the \textit{Solar Magnetic Activity Research Telescope} (SMART/SDDI).
We determined the physical quantities of the ejecta by cloud model fitting to the H$\alpha$  spectrum.
We determined flare energy by differential emission measure analysis using \textit{Atmospheric Imaging Assembly} onboard \textit{Solar Dynamics Observatory} (SDO/AIA) for small-scale flares and by estimating the bolometric energy for large-scale flares.
As a result, we found that the ejection mass $M$ and the total flare energy $E_{\mathrm{tot}}$ follow a relation of $M\propto E_{\mathrm{tot}}^{2/3}$.
We show that the scaling law  derived from a simple physical model explains the solar and stellar observations with a coronal magnetic field strength as a free parameter.
We also found that the kinetic energy and velocity of the ejecta correlate with the flare energy.
These results suggest a common mechanism driven by magnetic fields to cause cold plasma ejections with flares on the Sun and stars.

\end{abstract}



\section{Introduction} \label{sec:intro}

Filament eruptions are phenomena in which low-temperature plasma ($\sim10^4\,\mathrm{K}$) in the solar corona erupts into the interplanetary space, and they are often accompanied by flares and coronal mass ejections (CMEs) \citep{2014LRSP...11....1P}.
In the standard flare model, filament eruptions are considered to be the trigger of flares \citep{1995ApJ...451L..83S}.
Filaments can be broadly classified into active region filaments, quiescent filaments, and intermediate filaments \citep{2010SSRv..151..333M}, all of which can be erupted.
After the filament eruption, post-flare loops are formed in the case of active region filament eruptions, and giant arcades are formed in the case of quiescent filament eruptions. 
Post-flare loops and giant arcades have different X-ray intensities and spatial scales, but their morphologies are similar.
Therefore, both of these phenomena are considered to be explained by the standard flare model \citep{2011LRSP....8....6S}.

The typical length of an ejected filament ranges from $10^4$ to $10^5\,\mathrm{km}$, but signatures of smaller/larger spatial scale ejections of cold plasma with flares have also been observed.
Small plasma ejections with chromospheric temperatures over $10^3$ to $10^4\,\mathrm{km}$ (called minifilaments) occur in the Sun and are considered miniature versions of filament eruptions.
These small phenomena were first discovered in the quiet region in the 1970s \citep{1977ApJ...218..286M,1979SoPh...61..283L}.
\citet{1986NASCP2442..369H} performed the first comprehensive study by H$\alpha$ observation.
Subsequent observations showed that they were accompanied by extreme ultraviolet (EUV) small flares \citep{2004ApJ...616..578S,2008Ap&SS.318..141R}.
These early studies also revealed that minifilament eruptions are associated with magnetic flux cancellation.
In addition, minifilament eruptions are often accompanied by coronal waves and mini-CMEs \citep{2009A&A...495..319I,2010ApJ...709..369P,2010ApJ...710.1480S}.
Several studies reported that minifilament eruption contributes to the EUV jet formation with flux cancellation \citep[e.g.,][]{2015Natur.523..437S}.
These previous studies suggest that minifilament eruptions have common properties with filament eruptions and flares, and recent observations with high spatial resolution have supported the same conclusion \citep{2020ApJ...898..144K,2020ApJ...902....8C,2021ApJ...914L..35J}.
Cold plasma ejections have also been found with campfires, which are smaller EUV brightenings recently discovered by the Solar Orbiter/Extreme Ultraviolet Imager \citep{2021ApJ...921L..20P,2021A&A...656L...4B}.
Considering the scale-free self-similar characteristics of magnetohydrodynamics, we can expect similar phenomena at smaller scales \citep{2020PASJ...72...75K}.
Even in the case of stellar flares occurring on M-type stars, which are sometimes much stronger than solar flares, blue asymmetry has been found in spectral profiles of the chromospheric lines \citep{1990A&A...238..249H,2016A&A...590A..11V,2019A&A...623A..49V,2018PASJ...70...62H,2019ApJ...877..105M,2021PASJ...73...44M}.
A plausible interpretation is that the blue asymmetry represents  cold plasma ejections associated with stellar flares.
Also, a blue shift in the H$\alpha$ absorption line has been found with a superflare in G-type stars\citep{2021NatAs...6..241N}.
Because of its similarity to filament eruptions on the Sun, it is reasonable to interpret this as large cold plasma ejections associated with a stellar flare.

Cold plasma ejections occur over a wide range of flare energy spanning more than ten orders of magnitude ($10^{25}-10^{35}\,\mathrm{erg}$).
In contrast, the relationships between their physical parameters (such as ejection mass, kinetic energy, and velocity) and flare energy have not been studied quantitatively as a common framework.
The ejection mass and flare energy have been estimated by spectral observation of the stellar chromosphere in the stellar flare studies \citep{2019ApJ...877..105M,2021PASJ...73...44M,2021NatAs...6..241N}.
These stellar studies report that the stellar flare energy and the mass of associated ejecta could be interpreted as an extension of the solar CMEs.
However, there should be a significant gap in the temperature and observable height of the target between the solar CMEs observation using coronagraphs and the observation of cold plasma ejections using chromospheric lines.
Therefore, the cold plasma ejections associated with stellar flares are strictly speaking in a different framework than the solar CMEs.
The stellar cold plasma ejections should be compared to the solar cold plasma ejections.
\citet{2021NatAs...6..241N} has compared solar filament eruptions with stellar cases, but the number of samples is only five \citep{1987JApA....8..295J,1999spro.proc..367O,2015ApJ...804..147C,2021NatAs...6..241N}.
In addition, to our knowledge, the flare energy and the ejection mass of events smaller than $10^{27}\,\mathrm{erg}$ have never been quantitatively studied. To investigate the correlation between the physical quantities of the cold ejecta and the flare energy over a wide energy range, a larger sample of flares, including those below $10^{27}\,\mathrm{erg}$, needs to be examined.

In this study, we performed a statistical spectral analysis of small mass ejections associated with small flares  in the quiet region of the Sun (energy $10^{25}-10^{27}\,\mathrm{erg}$) and filament eruptions associated with solar flares (energy $10^{27}-10^{29}\,\mathrm{erg}$) using the \textit{Solar Dynamics Doppler Imager} \citep[SDDI:][]{2017SoPh..292...63I} on the \textit{Solar Magnetic Activity Research Telescope} \citep[SMART:][]{2004SPIE.5492..958U}.
By adding small events to the sample of large events, including the stellar flares, this study aims to investigate the correlation between the flare energy and the physical parameters of cold plasma ejecta over the energy range of $10^{25}-10^{35}\,\mathrm{erg}$.
We analyzed small-scale ($10^{25}-10^{27}\,\mathrm{erg}$) phenomena that have common properties with minifilament eruption, such as line-of-sight velocity, lifetime, and appearance with small flares (=EUV brightening) (Figures \ref{fig:event} and \ref{fig:time_dev}).
Because shapes of the small ejecta are ambiguous due to the insufficient spatial resolution of H$\alpha$ observation, we refer to the small phenomenon to be analyzed in this article as ‘‘small mass ejections associated with small flares in the quiet region'' rather than ‘‘mini filament eruption."

We also constructed a theoretical scaling law between the total flare energy\footnote{This paper use the term ``total flare energy'' as an alternative to dissipated magnetic energy in magnetic reconnection. Dissipated magnetic energy may be best suited to define the flare energy over a wide range of energies; however, it is difficult to estimate directly from observations. On the other hand, the kinetic energy and the energies related to the flare radiation (bolometric, non-thermal and thermal) can be estimated from observations. In addition, the dissipated magnetic energy is mainly accounted for by kinetic and radiation-related energies \citep{2012ApJ...759...71E}. Therefore, we define the ``total flare energy'' as the sum of the kinetic energy of the cold plasma and the energy related to the flare radiation (see Section \ref{sec:obs}).}  and the mass of ejected filament and compared it with our observation results.
Using this scaling law, we attempt to understand cold plasma ejections associated with flares of various scales in a unified way.

\section{Observation and Data Analysis} \label{sec:obs}

We used SMART/SDDI to detect cold plasma ejections.
SDDI at Hida Observatory of Kyoto University takes the solar full-disk images at 73 wavelengths from H$\alpha-9.0\,$\AA\, to H$\alpha+9.0\,$\AA\, with a constant wavelength step of 0.25 \AA.
The temporal resolution and the pixel size are 12 s and 1.23 arcsec, respectively.
The data were processed for dark and flat field correction.

\subsection{Small Mass Ejections Associated with Small Flares in the Quiet Region} \label{sec:subobs}

\begin{figure}
\epsscale{0.8}
\plotone{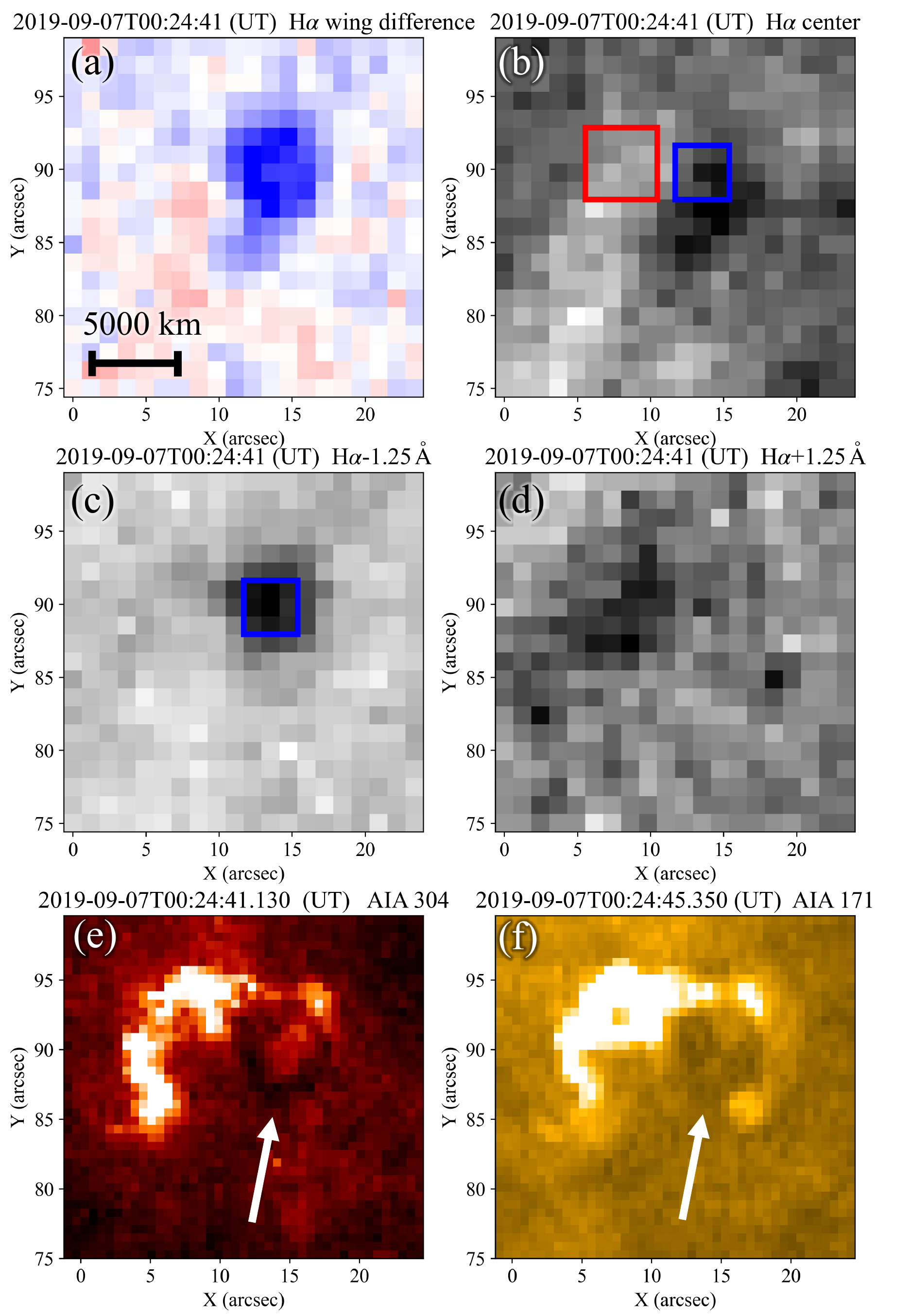}
\caption{
A typical example of a small mass ejection in the quiet region with a small flare.
(a) H$\alpha$ wing difference image.
This image is created by subtracting the counts from H$\alpha +1.25$ \AA\, to $+2.0$ \AA\, from the counts from H$\alpha -1.25$ \AA\, to $-2.0$ \AA.
For the colors in this figure, blue and red represent dark and bright features corresponding to the blue and red shifts.
(b) H$\alpha$ line center intensity.
The red and blue squares represent the area where the H$\alpha$'s light curve was calculated and the temporal variation of the line-of-sight velocity were calculated in Figure \ref{fig:time_dev}, respectively.
(c), (d) Intensities at H$\alpha\pm1.25$ \AA.
The blue square is the same with that in panel (b).
(e), (f) AIA 304 \AA\, and 171 \AA\, intensities.
The white arrows indicate mass ejections.
}\label{fig:event}
\end{figure}

\begin{figure}
\epsscale{0.8}
\plotone{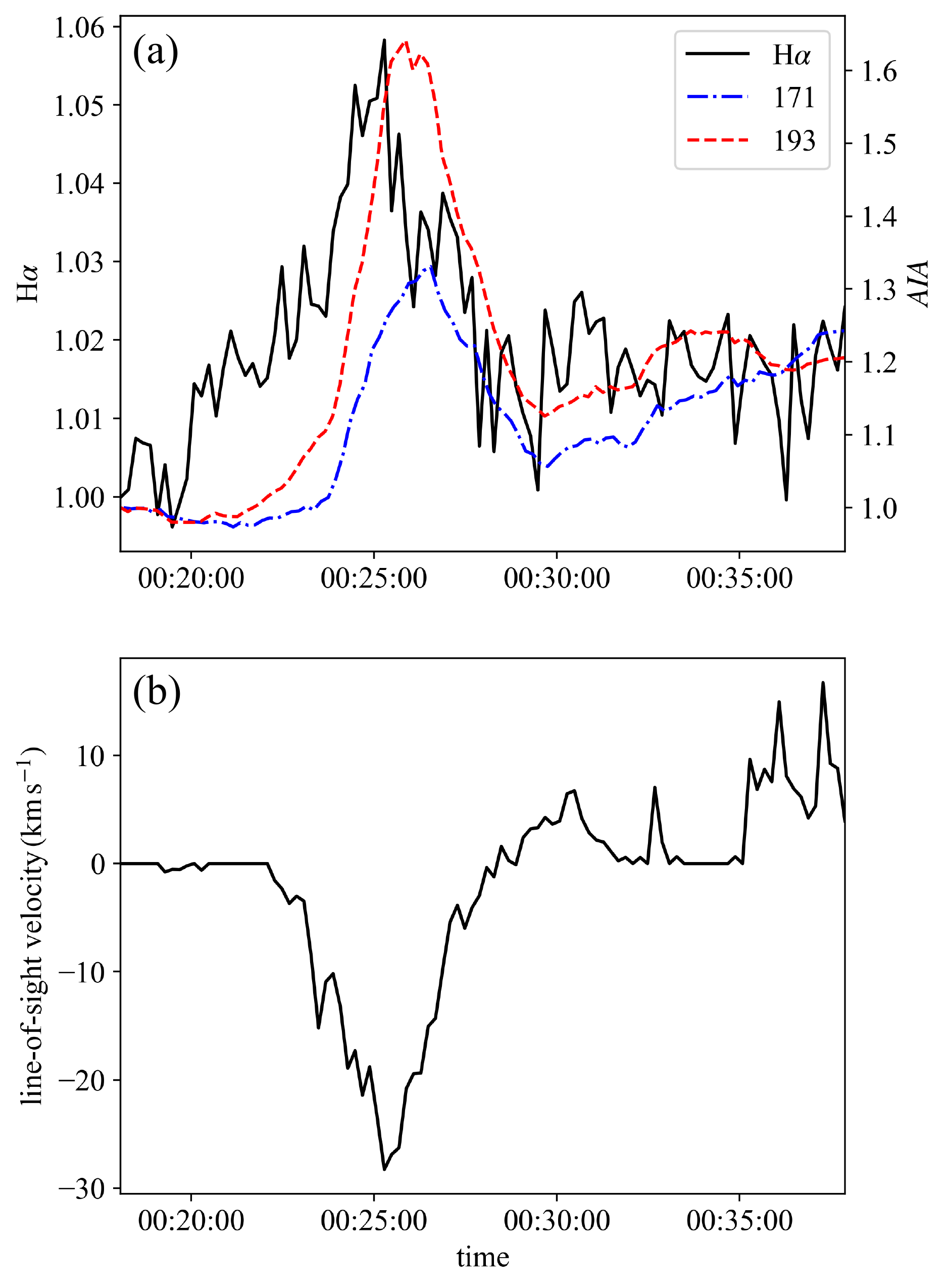}
\caption{
(a) Light curves of H$\alpha$ center (solid black), AIA 171 \AA\ (dash-dotted blue), and 193 \AA\  (dashed red).
H$\alpha$ brightness is calculated as the average over the red square region in Figure \ref{fig:event}b, and 171 \AA\, AIA 193 \AA\, are calculated as the average over the entire field of view in Figure \ref{fig:event}f.
Each light curve is normalized by its initial brightness at 00:18 (UT).
(b) Temporal variation of the line-of-sight velocity of the small mass ejection.
The line-of-sight velocity is calculated by averaging those obtained by the cloud model fitting in the blue squares in Figure \ref{fig:event}b.
A negative value represents a blue shift, and a positive value represents a red shift.
}\label{fig:time_dev}
\end{figure}

\begin{figure}
\epsscale{0.8}
\plotone{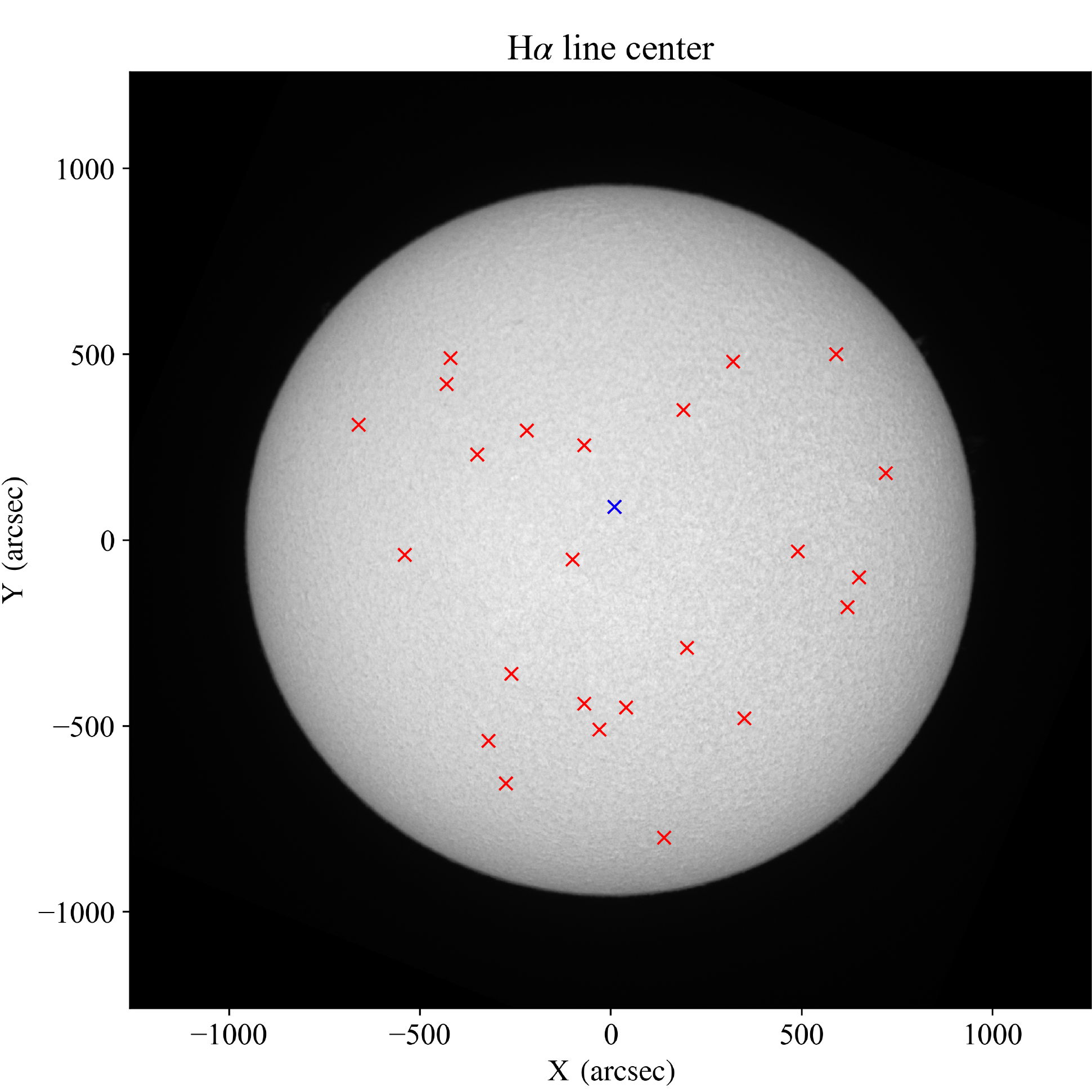}
\caption{
Solar full-disk image observed by SMART/SDDI at the H$\alpha$ line center on September 7, 2019.
The red crosses indicate the locations of the small mass ejections used in the analysis.
The blue cross indicates the event shown in Figures \ref{fig:event} and \ref{fig:time_dev}.
}\label{fig:event_list}
\end{figure}

We used the SDDI data taken from 21:56:17 UT on September 6, 2019 to 8:32:40 UT on September 7, 2019 to detect the small mass ejections in the quiet region.
In this case, we used 25 wavelengths covering from H$\alpha-3.0\,$\AA\ to H$\alpha+3.0\,$\AA\ because this range is sufficient to capture our targets.
On this day, neither sunspots, large plage regions, nor large filaments were present on the solar disk (Figure \ref{fig:event_list}).
Although the Sun was tranquil, tiny spatial dimming was frequently observed in the H$\alpha$ blue wing.

We analyzed 25 events that satisfied the following conditions.
 \begin{itemize}
  \item They showed clear dark features in “H$\alpha$ wing difference images", which corresponded to the blue shift due to the fast mass ejections.  
  We made these wing difference images by subtracting ‘‘the average of the counts from H$\alpha +1.25$ \AA\, to $+2.0$ \AA'' from ‘‘the average of the counts from H$\alpha -1.25$ \AA\, to $-2.0$ \AA'' at each time.
  These dark structures include pixels where the difference counts are larger than 4$\sigma$ ($\sigma=$ the standard deviation calculated from the surrounding pixels).
  \item They were located within approximately $60^{\circ}$ from the disk center.
  \item They were associated with EUV brightening in images taken by the \textit{Atmospheric Imaging Assembly} (AIA)  onboard \textit{Solar Dynamics Observatory} (SDO)
  \citep{2012SoPh..275....3P,2012SoPh..275...17L}.
  As described in detail later, we checked for the presence of areas at least three times brighter than the surroundings as a criterion for brightness.
 \end{itemize}
Locations of the selected 25 events are shown in Figure \ref{fig:event_list}.
Figures \ref{fig:event} and \ref{fig:time_dev} show a typical example of the event and its time development.

To estimate the kinetic energy and mass of the ejecting plasma, we performed cloud model fitting \citep{1964PhDT........83B,1988A&A...203..162M} to determine the line-of-sight velocity $v_l$ (in units of $\mathrm{cm}\,\mathrm{s}^{-1}$), optical thickness of the ejecting plasma $\tau_0$, source function $S$, and Doppler width $\Delta\lambda_D$ (in units of \AA).
Using the fitting results, we counted the number of pixels $A_{\mathrm{ejecta}}$ where the blue shift was greater than $10\,\mathrm{km}\,\mathrm{s}^{-1}$ and defined it as the area of the ejections.

\begin{figure}
\epsscale{1.0}
\plotone{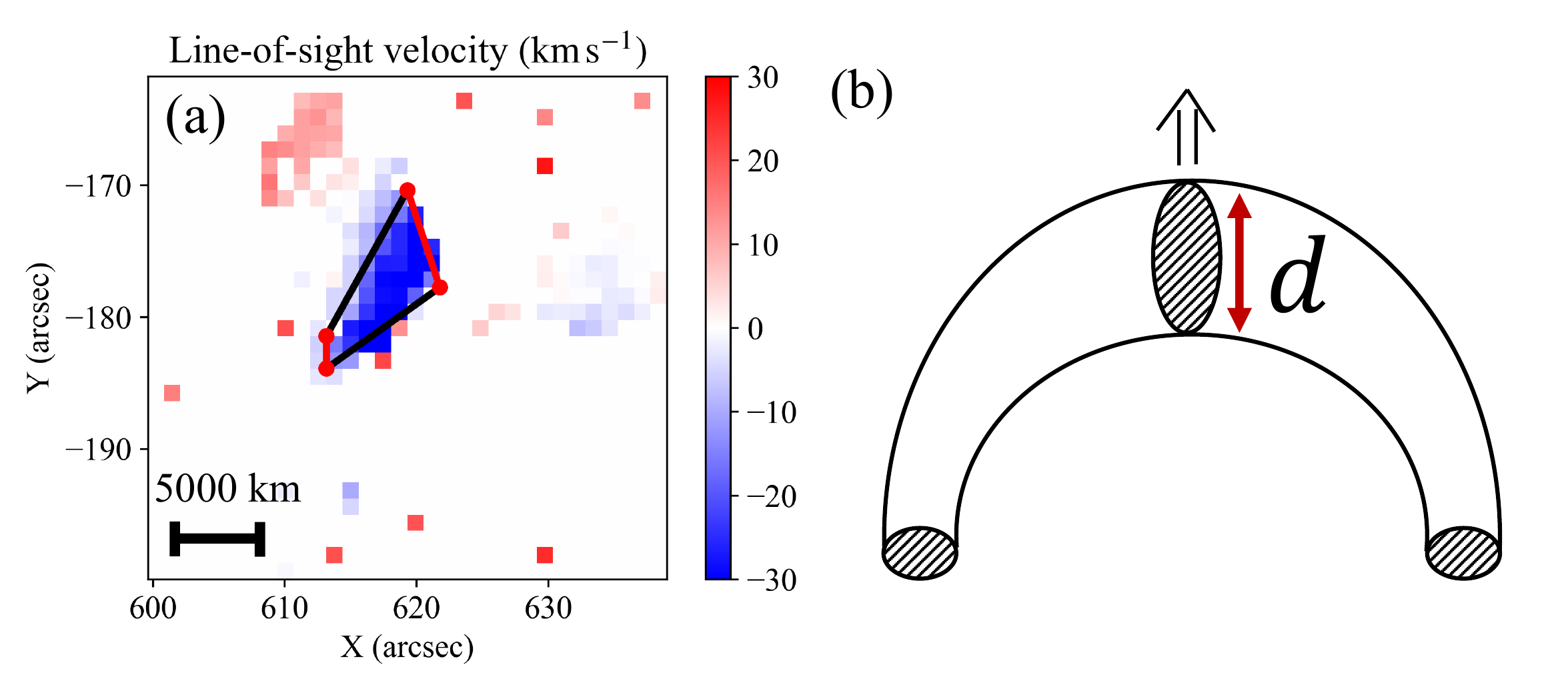}
\caption{
Diagrams showing how the length $L$, width, and line-of-sight length $d$ of cold ejecta are defined.
(a): Typical example for the approximation of cold ejecta as a quadrangle.
The color bar indicates the line-of-sight velocity $(\mathrm{km}\,\mathrm{s^{-1}})$.
The red and black lines indicate the opposite sides corresponding to width and length $L$, respectively.
(b): Schematic diagram of the eruption assumed in this study. $d$ is the line-of-sight length.
}\label{fig:width_estimate}
\end{figure}

We estimated the physical quantities of each event when the sum of $v_l^2$ in region $A_{\mathrm{ejecta}}$  reached the maximum.
The sum of $v_l^2$ can be regarded as a measure of the kinetic energy of the ejection if the mass column density is uniform over the ejection.
To estimate the hydrogen density $n_H$ (in units of $\mathrm{cm}^{-3}$), the second level hydrogen density $n_2$\footnote{$n_2$ includes the effect of the filling factor. This is because $\tau_0$ and $\Delta\lambda_D$ are determined from the spectrum whose signal is weakened by the filling factor.}  (in units of $\mathrm{cm}^{-3}$) was obtained from the results of the cloud model fitting as follows \citep{1997A&A...324.1183T}:
\begin{equation}
n_2 = 7.26\times10^{12} \frac{\tau_0 \Delta\lambda_D}{d}\,\mathrm{cm}^{-3}, \label{N2}
\end{equation}
where $d$ is the line-of-sight length of the ejecta (in units of $\mathrm{cm}$).
We approximated $A_{\mathrm{ejecta}}$ as a quadrilateral by selecting the four endpoints, as shown in Figure \ref{fig:width_estimate}a.
We calculated the average of the opposite side lengths of the quadrangle, respectively, and defined the shorter and longer sides as the width and length $L$ of the ejecta.
Assuming that the cold plasma erupts in a cylinder-like structure (Figure \ref{fig:width_estimate}b), we defined the line-of-sight length $d$ as equal to the width.
We estimated $n_H$, column density $M_{\mathrm{col}}$  ($\mathrm{g}\,\mathrm{cm}^{-2}$), and mass density $\rho$ ($\mathrm{g}\,\mathrm{cm}^{-3}$) from the $n_2$ values in each pixel as follows \citep{1997A&A...324.1183T}:
\begin{eqnarray}
&n_H & =  5.01\times 10^{8} \sqrt{n_2},  \label{NH}\\
&M_{\mathrm{col}}& = (n_H  m_H + 0.0851 n_H \times 3.97m_H)d, \label{Mcol}\\
&\rho& = \frac{M_{col}}{d}, \label{rho}
\end{eqnarray}
where $m_H= 1.67\times10^{-24}\,\mathrm{g}$ is the mass of the hydrogen atom.
Using equation (\ref{Mcol}), we obtained the mass $M$ ($\mathrm{g}$) and kinetic energy $E_k$ ($\mathrm{erg}$) of the ejecting plasma by summing over the pixels in region $A_{\mathrm{filament}}$;
\begin{eqnarray}
&M& = \sum_{A}M_{\mathrm{col}}\times(\text{SDDI pixel size})^2, \label{MM} \\
&E_k& = \sum_{A}\frac{1}{2}\rho v_{l}^{2}\times(\text{SDDI pixel size})^2\times d, \label{Ek}
\end{eqnarray}
where SDDI pixel size $= 1.23\,\mathrm{arcsec}= 8.99 \times 10^7\,\mathrm{cm}$.

To determine the total flare energy, we estimated the thermal energy of the small flare by the differential emission measure (DEM) from the AIA 6 channels. 
The analysis was performed for the dataset at the peak time of AIA 193 \AA\ light curve.
We extracted the areas where the intensity of AIA 171 \AA\ and 193 \AA\ was more than five times brighter than the surrounding area and where AIA 131 \AA\ and 211 \AA\  were more than seven times brighter than the surrounding area.
These areas were combined to determine the region for DEM analysis.
In three events,  the intensity of the surrounding area was high, or the brightening was low, and we could not extract the area for DEM analysis by the above method.
In these cases, we determined the area for DEM analysis with the threshold of  AIA 171\AA\ and 193 \AA\ intensity as three times, and AIA 131 \AA\ and 211 \AA\ intensity as five times.
We defined the spatial scale of the flare $L_{\mathrm{flare}}$ (in units of $\mathrm{cm}$) by taking the square root of the area where the DEM analysis was performed.

We performed the DEM analysis using the method of \citet{2012A&A...539A.146H}.
We used AIA level 1.5 data generated by $\mathsf{sunpy}$'s $\mathsf{aiapy.calibrate}$ as input values for the DEM analysis.
We determined the DEM in the range $5.5\le\log T\le6.7$ with a step of $d\log T = 0.05$.
After obtaining the DEM, we calculated the emission measure $EM(\mathrm{cm}^{-5})$, DEM-weighted temperature $T_{DEM}$ ($\mathrm{K}$), and electron density $n_e$ ($\mathrm{cm}^{-3}$) at each pixel using the following relations:
\begin{eqnarray}
EM &=& \int DEM(T)dT   \label{EM}\\
T_{DEM}&=&\frac{\int DEM(T)\times TdT}{\int DEM(T)dT}  \label{TDEM}\\
n_e &=& \sqrt{\frac{EM}{L_{\mathrm{flare}}}}.  \label{ne}
\end{eqnarray}
We used these values to obtain the thermal energy of the small flare $E_{th}$ (in units of $\mathrm{erg}$) by summing over all pixels for which the DEM analysis was performed.
\begin{equation}
E_{th} = \sum 3n_{e}k_{B}T_{DEM}, \label{Eth}
\end{equation}
where $k_B$ is Boltzmann constant.

We determined the total flare energy $E_{\mathrm{tot}}$ ($\mathrm{erg}$) as the sum of the kinetic energy of the cold ejection and the thermal energy for small flares cases.
Previous studies for the energy partition of flares have suggested that bolometric energy is a good alternative for energies related to flare radiation \citep{2012ApJ...759...71E}.
However, we could not estimate the bolometric energy because all the small flares analyzed in this study are sub-A class flares. 
On the other hand, \citet{2020A&A...644A.172W} have proposed that the contribution of non-thermal energy decreases as the flare energy decreases.
 We, therefore, followed the hypothesis of \citet{2020A&A...644A.172W} and used thermal energy as a bolometric energy substitute, neglecting the non-thermal energy contribution.

\begin{figure*}
\epsscale{1.0}
\plotone{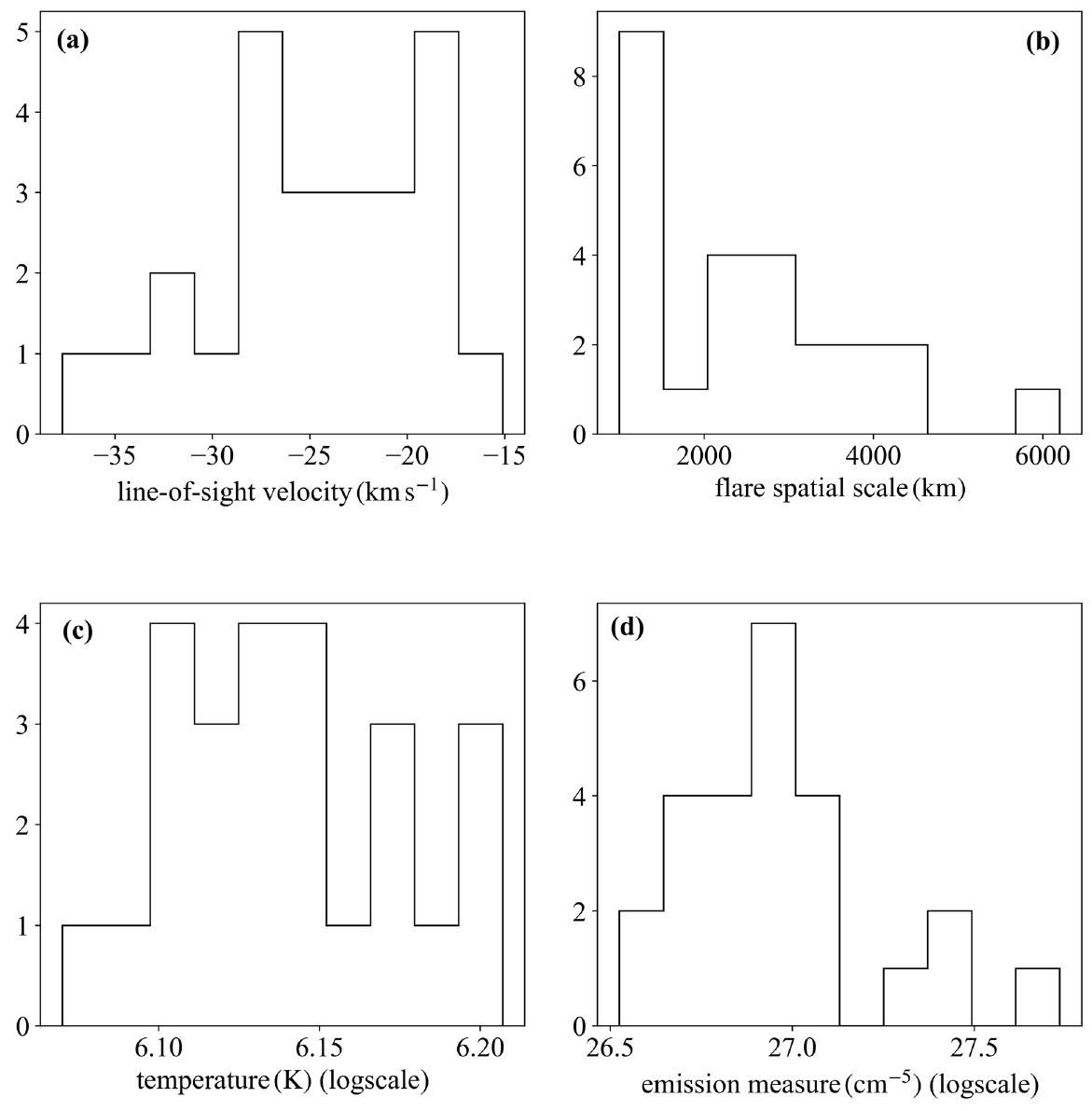}
\caption{
Histograms of the physical parameters of 25 events analyzed in this study.
(a) Line-of-sight velocity of the cold plasma ejection obtained by the cloud model.
(b)-(d) Spatial scale, temperature, and EM for small-scale flares.
}\label{fig:para}
\end{figure*}

Distributions of the physical quantities thus obtained are shown in Figure \ref{fig:para}a-\ref{fig:para}d.
The result of the line-of-sight velocities in this study (Figure \ref{fig:para}a) is larger than those by \citet{2020ApJ...898..144K}, which reported about $10\,\mathrm{km}\,\mathrm{s^{-1}}$, also based on the cloud model.
This is probably because we used H$\alpha$ wing difference images from $\pm1.25$  \AA\,  to $\pm2.0$ \AA\, in our event selection, which is sensitive to faster velocity movements.
Spatial scale, temperature, and EM for small-scale flares (Figures \ref{fig:para}b, \ref{fig:para}c, and \ref{fig:para}d, respectively) are comparable to or slightly greater than those studied for the larger “campfires" \citep{2021A&A...656L...4B}.
Note that we have obtained the temperature and EM for each event by averaging the physical quantities for pixels where the DEM analysis was performed.
We can confirm from Figure \ref{fig:para} that we are indeed analyzing a small flare associated with a cold plasma ejection.

\subsection{Filament Eruptions with Solar Flares} \label{sec:subSeki}

To perform the similar study for large filament eruptions, we used the SMART/SDDI filament disappearance catalog\footnote{https://www.kwasan.kyoto-u.ac.jp/observation/event/sddi-catalogue/} 
\citep{2019SunGe..14...95S} as the data source.
From this catalog, we selected events near the disk center that were accompanied by flares.
Under this criterion, we selected six active region filaments and four intermediate filaments; in total, ten cases were analyzed in this study.

We estimated the kinetic energy of the filament eruption using the cloud model fitting in the same way as the small mass ejections.
The analysis was performed for the period when the radial velocities of the eruption got maximum.
We estimated the bolometric energy by assuming that the GOES flare index is proportional to the bolometric energy and that the bolometric energy of the M1.0-class flare corresponds to $10^{30}\,\mathrm{erg}$ \citep{2013PASJ...65...49S,2021NatAs...6..241N}.
As a result, we obtained the total flare energy as the sum of the kinetic energy of the filament eruption and the bolometric flare energy.

\section{Results} \label{sec:result}

\begin{figure*}[t!]
\epsscale{0.75}
\plotone{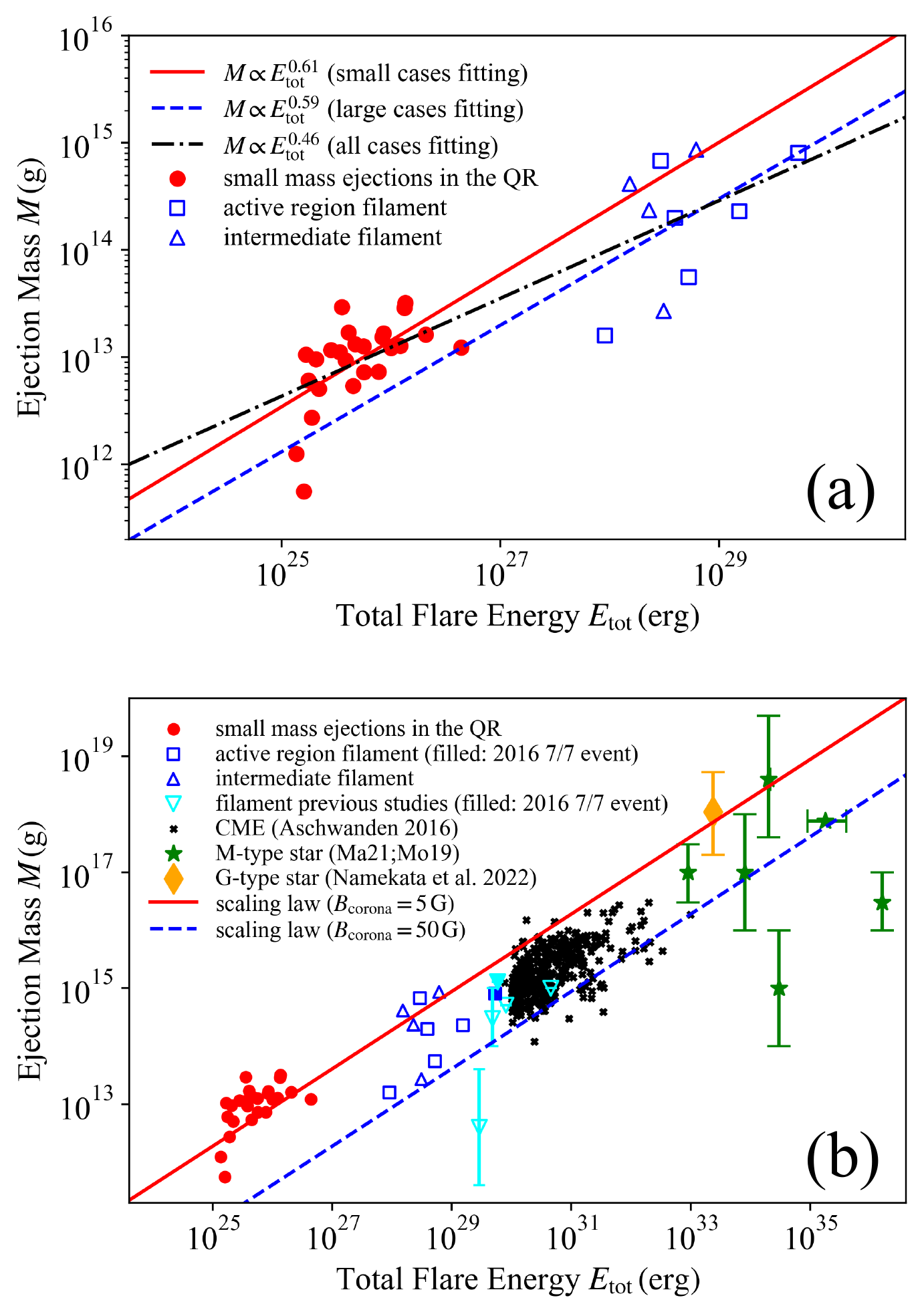}
\caption{
(a) Relationship between total flare energy and ejected mass for events analyzed in this study.
The red circles, blue squares, and blue triangles represent small mass ejections in the quiet region, the active region filament, and the intermediate filament eruptions, respectively.
The solid red and dashed blue lines show the results of fitting for red symbols and blue symbols, respectively.
The dash-dotted black line shows the fitting of all events in this figure.
(b) Comparison of the present study with previous studies.
The light blue triangles represent filament eruptions in previous studies \citep{1987JApA....8..295J,1999spro.proc..367O,2015ApJ...804..147C,2021NatAs...6..241N}.
The filled blue square and light blue triangle represent the same event, the July 7, 2016 filament eruption analyzed in \citet{2021NatAs...6..241N}.
The black crosses represent CMEs \citep{2016ApJ...831..105A}.
The green stars and orange diamond represent signs of cold plasma ejections with stellar flares on M-type \citep[shown Mo19 and Ma21 in the figure]{2019ApJ...877..105M,2021PASJ...73...44M} and G-type stars \citep{2021NatAs...6..241N}, respectively. 
The solid red line and the dashed blue line show the cases of $B_{\mathrm{corona}}=5\,\mathrm{G}$ and  $B_{\mathrm{corona}}=50\,\mathrm{G}$ in the scaling law  (\ref{MvsE_2}), respectively.
We assume the energy conversion rate $f = 0.1$ in equation (\ref{MvsE_2}) in both lines.
}\label{fig:all}
\end{figure*}

\begin{figure*}[t!]
\begin{center}
\epsscale{0.8}
\plotone{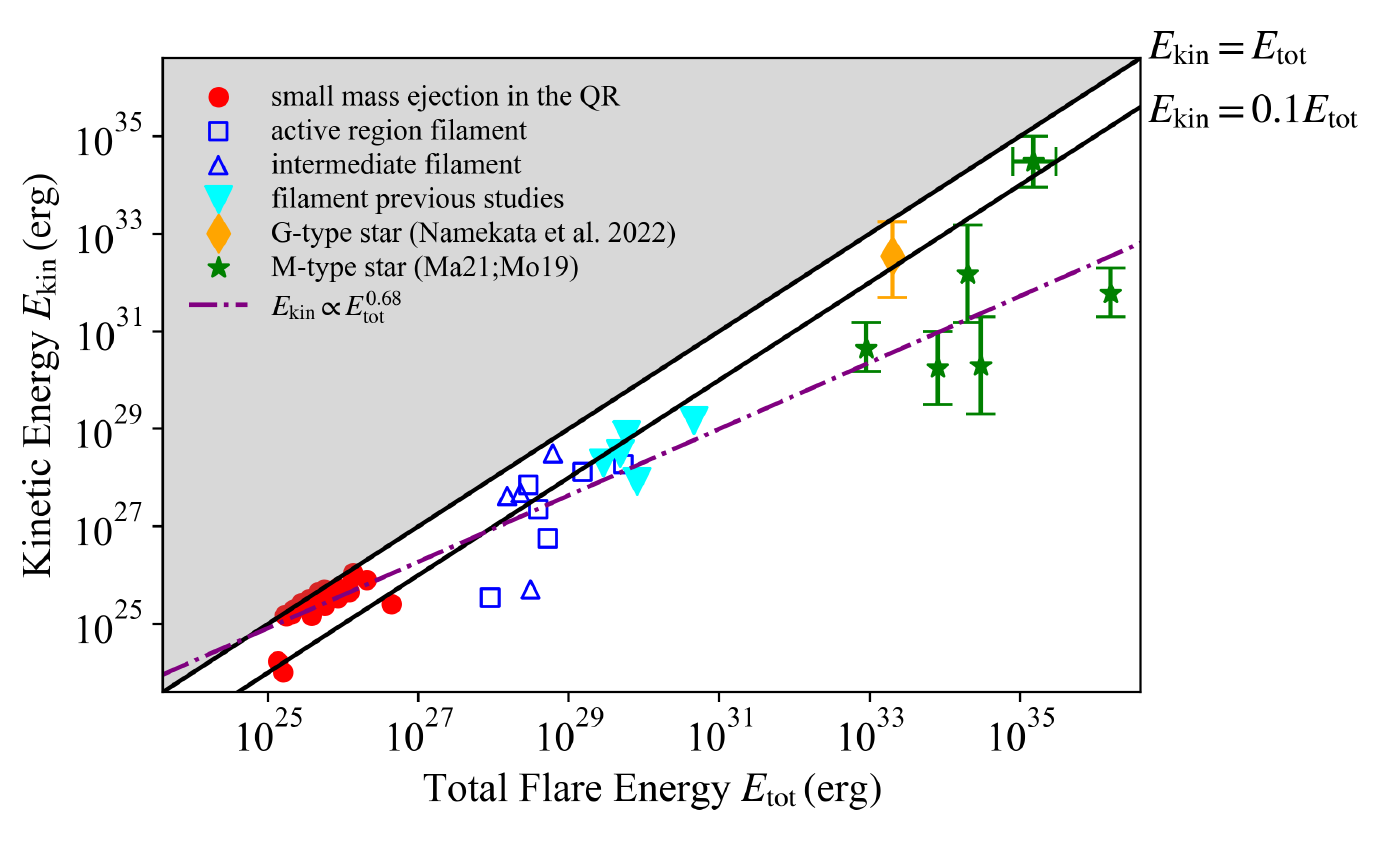}
\caption{
Same as Figure \ref{fig:all} but for the kinetic energy.
The solid black lines shows $E_{\mathrm{kin}}=E_{\mathrm{tot}}$ and $E_{\mathrm{kin}}=0.1E_{\mathrm{tot}}$, respectively.
The dash-dotted purple line shows the fitting of all events in Figure \ref{fig:all}a.
The gray region shows $E_{\mathrm{kin}}>E_{\mathrm{tot}}$.
}\label{fig:all_kine}
\end{center}
\end{figure*}

\begin{figure*}[t!]
\begin{center}
\epsscale{0.75}
\plotone{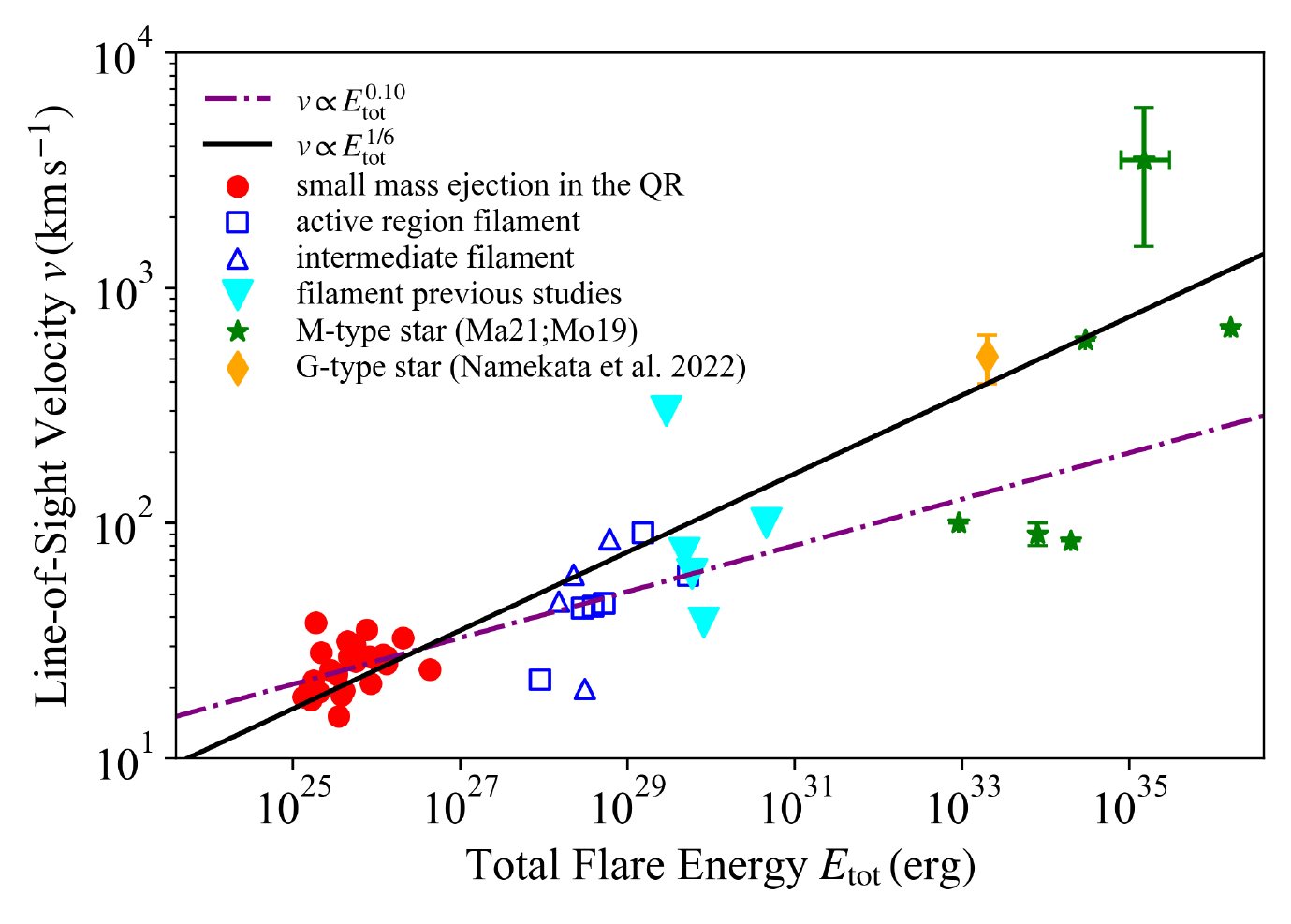}
\caption{
Same as Figure \ref{fig:all} but for the line-of-sight velocity.
The dash-dotted purple line shows the fitting of all events in Figure \ref{fig:all}a.
The solid black line show $v\propto E_{\mathrm{tot}}^{1/6}$ relation.
}\label{fig:all_vel}
\end{center}
\end{figure*}

Figure \ref{fig:all}a shows the relationship between total flare energy and ejected mass for the events analyzed in this study.
The correlation coefficients between the ejected mass and total flare energy for the small mass ejections in the quiet region $r_{\mathrm{small}}$ and filament eruptions $r_{\mathrm{large}}$ are $0.60$ and $0.49$, respectively.
The 95\% confidence intervals for the respective correlation coefficients are $0.27<r_{\mathrm{small}}<0.80$ and $-0.21<r_{\mathrm{large}}<0.85$.
These values confirm a positive correlation between total flare energy and ejection mass at least in small event cases.
In the large filament eruption cases, we cannot claim a positive correlation from only these values, but this may be due to the small sample number of 10.
The correlation coefficient of all events analyzed in this study $r_{\mathrm{all}}$ is $r_{\mathrm{all}}=0.85$ and its 95\% confidence intervals is $0.73<r_{\mathrm{all}}<0.92$, suggesting a strong correlation over a wide range of energies as expected.
Linear fitting of the logarithms of the ejected mass and flare energy results in $M\propto E_{\mathrm{tot}}^{0.61\pm0.25}$ for small mass ejections in the quiet region and $M\propto E_{\mathrm{tot}}^{0.59\pm0.22}$ for filament eruptions, where the uncertainties are given by assuming the error in mass by a factor of three.
For all events analyzed in this study, $M\propto E_{\mathrm{tot}}^{0.46\pm0.06}$ was obtained.
It is noticed that the power-law index was reduced compared to the individual cases.

In Figure \ref{fig:all}b, we include other examples of solar filament eruption \citep{1987JApA....8..295J,1999spro.proc..367O,2015ApJ...804..147C,2021NatAs...6..241N} and the blue shifts accompanied by the stellar flares interpreted as stellar filament eruptions \citep{2019ApJ...877..105M,2021PASJ...73...44M,2021NatAs...6..241N} with the present analysis.
We assumed $E_{\mathrm{tot}}=100E_{\mathrm{X}} + E_{\mathrm{kin}}$ in the M-type star cases \citep{2019ApJ...877..105M,2021PASJ...73...44M} and $E_{\mathrm{tot}}=E_{\mathrm{bol}} + E_{\mathrm{kin}}$ in the G-type star case \citep{2021NatAs...6..241N}, where $E_{\mathrm{X}}$ is the X-ray energy in the GOES $1-8\,$\AA\, band of the stellar flares estimated from their H$\alpha$ energy, $E_{\mathrm{kin}}$ is the kinetic energy of the ejecta, and $E_{\mathrm{bol}}$ is the bolometric energy of white-light flare.
We can see a strong correlation between total flare energy and ejection mass over a wide range of energies in Figure \ref{fig:all}b.
The correlation coefficient $r_{\mathrm{all\_previous}}$, which includes both previous studies and stellar cases, increased to $r_{\mathrm{all\_previous}}=0.90$ and its 95\% confidence intervals is $0.83<r_{\mathrm{all\_previous}}<0.95$.

In Figure \ref{fig:all}b, we include the mass-total flare energy relation for the solar CMEs \citep{2016ApJ...831..105A}.
The total flare energy associated with CMEs is estimated in the same way as in the case of solar filament eruptions.
As a result, CMEs and cold plasma ejections show similar trends.
This trend is consistent with the scaling law for CMEs $M_{\mathrm{CME}}\propto E_{\mathrm{tot}}^{2/3}$ derived by \citet{2016ApJ...833L...8T}.
Observational studies of solar CMEs using X-ray flux and fluence as the flare energy also report power-law indexes close to 2/3 (0.70: \citealt{2011SoPh..268..195A}, 0.59: \citealt{2013ApJ...764..170D}).

Figure \ref{fig:all_kine} shows the relationship between total flare energy and kinetic energy of ejection for events analyzed in this study.
From this figure, we can see that the kinetic energy accounts for a larger percentage of the total energy for small ejections in the quiet region, while it becomes smaller to about 10\% for large filament eruptions.
The correlation coefficient of all events analyzed in this study is $r = 0.87$ and its 95\% confidence intervals is $0.76 < r < 0.94$.
The linear fitting between the logarithm of $E_{\mathrm{kin}}$ and $E_{\mathrm{tot}}$ results $E_{\mathrm{kin}}\propto E_{\mathrm{tot}}^{0.68\pm0.12}$.
Here, the uncertainty of the power index is estimated by assuming an error of an order of magnitude in the kinetic energy.
This power-law index is smaller than the power ($=1.05$) between the flare X-ray fluence and the CME kinetic energy reported by \citet{2013ApJ...764..170D}.
Figure \ref{fig:all_kine} also includes the previous result of the filament eruption and stellar events.
We can see from this figure that the results of the M-type stars roughly coincide to the extension of the fitting results of the events analyzed in this study.
The correlation coefficient $r_{\mathrm{previous}}$ increases to $r_{\mathrm{previous}}=0.96$ and its 95\% confidence intervals is $0.92<r_{\mathrm{previous}}<0.98$.

Figure \ref{fig:all_vel} shows the relationship between total flare energy and line-of-sight velocity  analyzed in this study.
The correlation coefficient of all events analyzed in this study is $r = 0.73$ and its 95\% confidence intervals is $0.52 < r < 0.85$.
Linear fitting of logarithms of these quantities results in $v\propto E_{\mathrm{tot}}^{0.10\pm0.04}$, where the uncertainty is given for an error of factor two in the velocity.
\citet{2016ApJ...833L...8T} derived $v\propto E_{\mathrm{tot}}^{1/6}$ for CMEs by assuming $E_{\mathrm{kin}}\propto E_{\mathrm{tot}}$ and $M\propto E_{\mathrm{tot}}^{2/3}$.
Since the filament eruptions in Figure \ref{fig:all_kine} are roughly consistent with the $E_{\mathrm{kin}}=0.1E_{\mathrm{tot}}$ relation, we also show the $v\propto E_{\mathrm{tot}}^{1/6}$ relation in the figure.
Figure \ref{fig:all_vel} also includes the comparison with previous results.
The correlation coefficient $r_{\mathrm{previous}}$ increases to $r_{\mathrm{previous}}=0.86$ and its 95\% confidence intervals is $0.76<r_{\mathrm{previous}}<0.92$.
Although the dependence on flare energy is small, we can confirm a positive correlation between velocity and total flare energy from this figure.
We can also confirm that the $v\propto E_{\mathrm{tot}}^{1/6}$ relationship roughly corresponds to the upper limit of the velocity.

\section{Discussion and conclusion} \label{sec:discss}

\subsection{Mass-Total Flare Energy Relation} \label{sec:sub_mass}

\begin{figure*}
\gridline{\fig{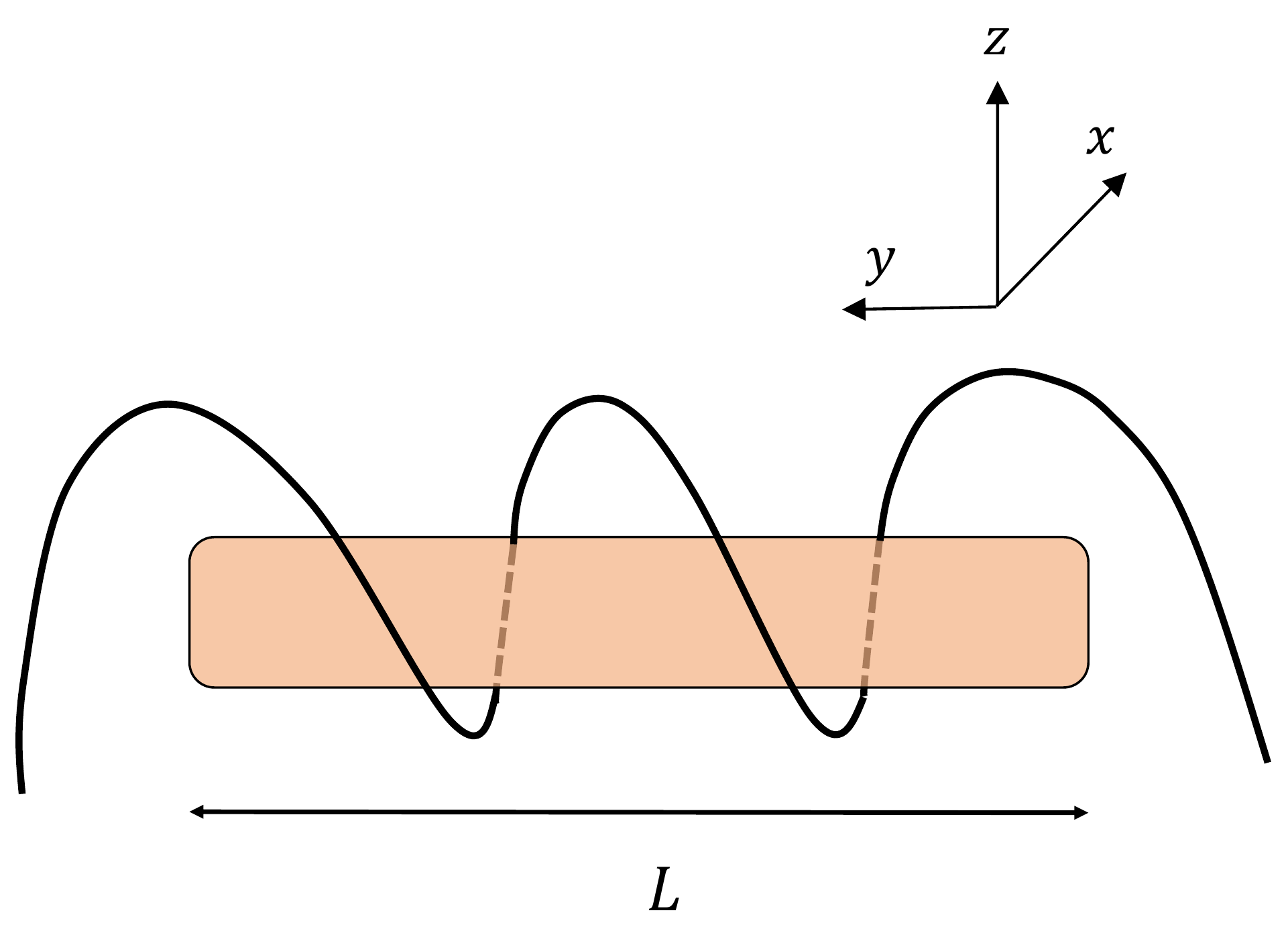}{0.4\textwidth}{(a)}
          \fig{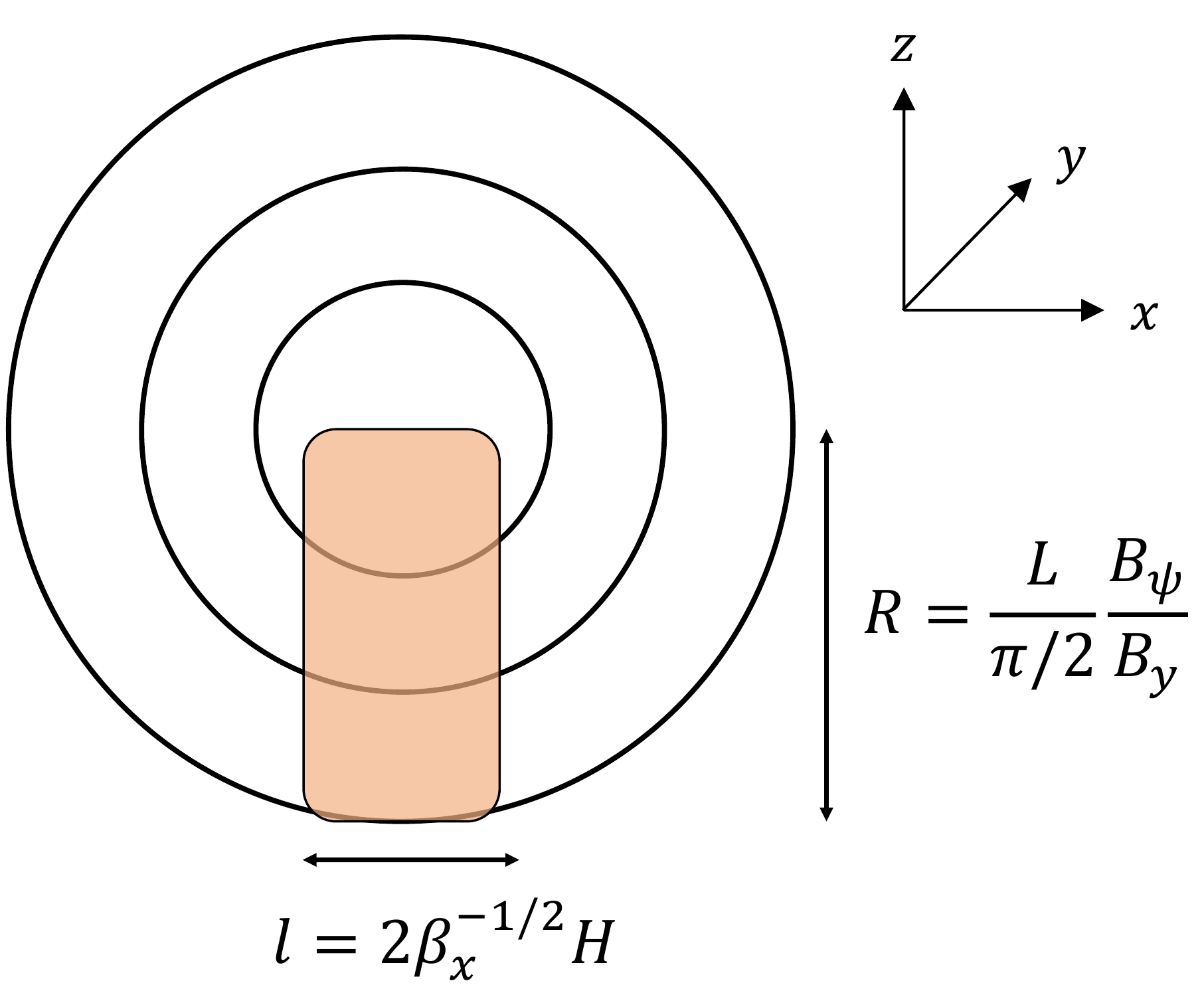}{0.4\textwidth}{(b)}}
\gridline{\fig{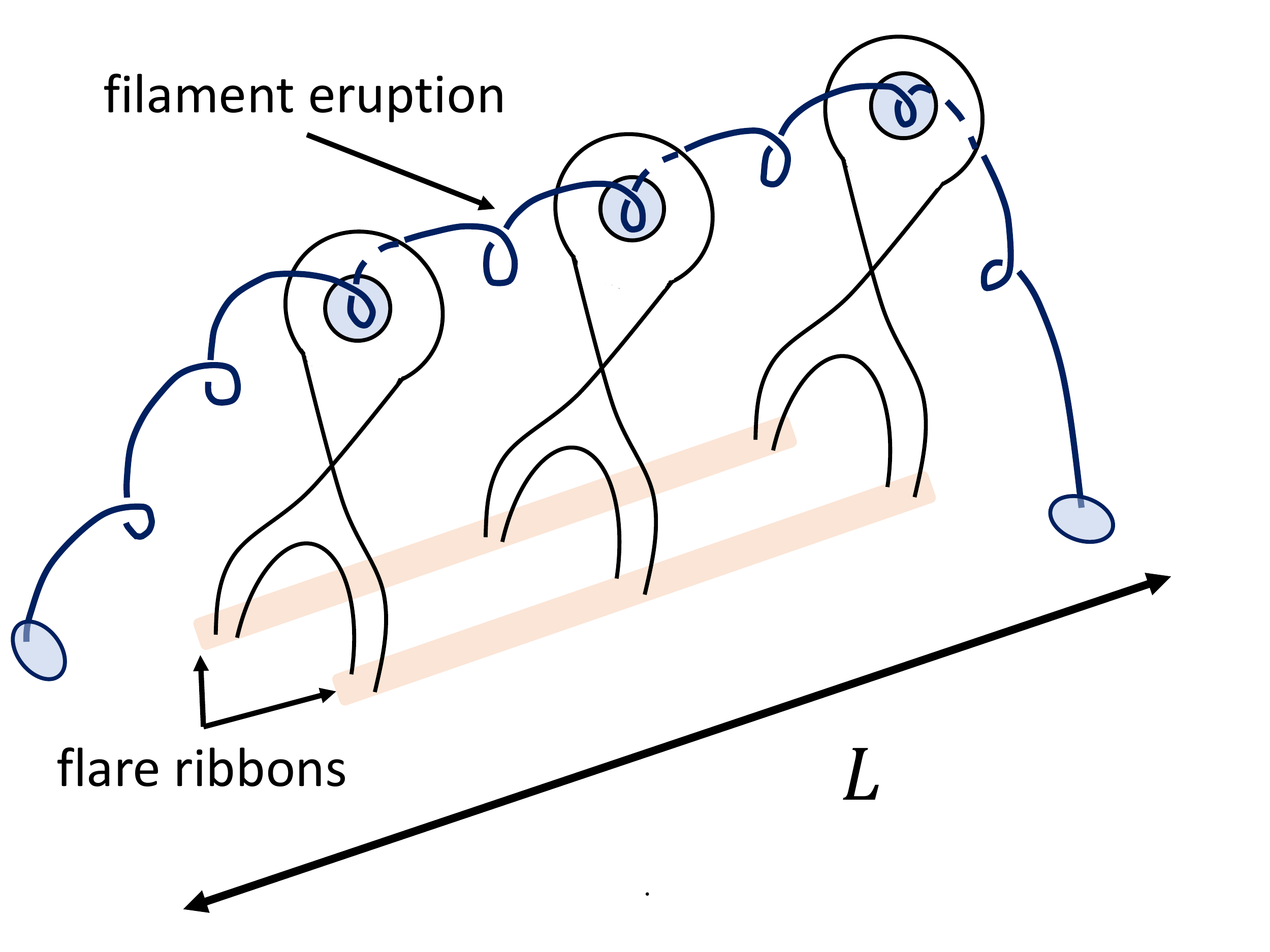}{0.8\textwidth}{(c)}}
\caption{
(a), (b) Geometric configuration of stably existing filaments.
For the filament height $R$, we show equation (\ref{KS_con_R}) assuming $a=4$.
(c) Relationship between the erupting filament and the flare.
}\label{fig:model}
\end{figure*}

\begin{figure}
\epsscale{0.8}
\plotone{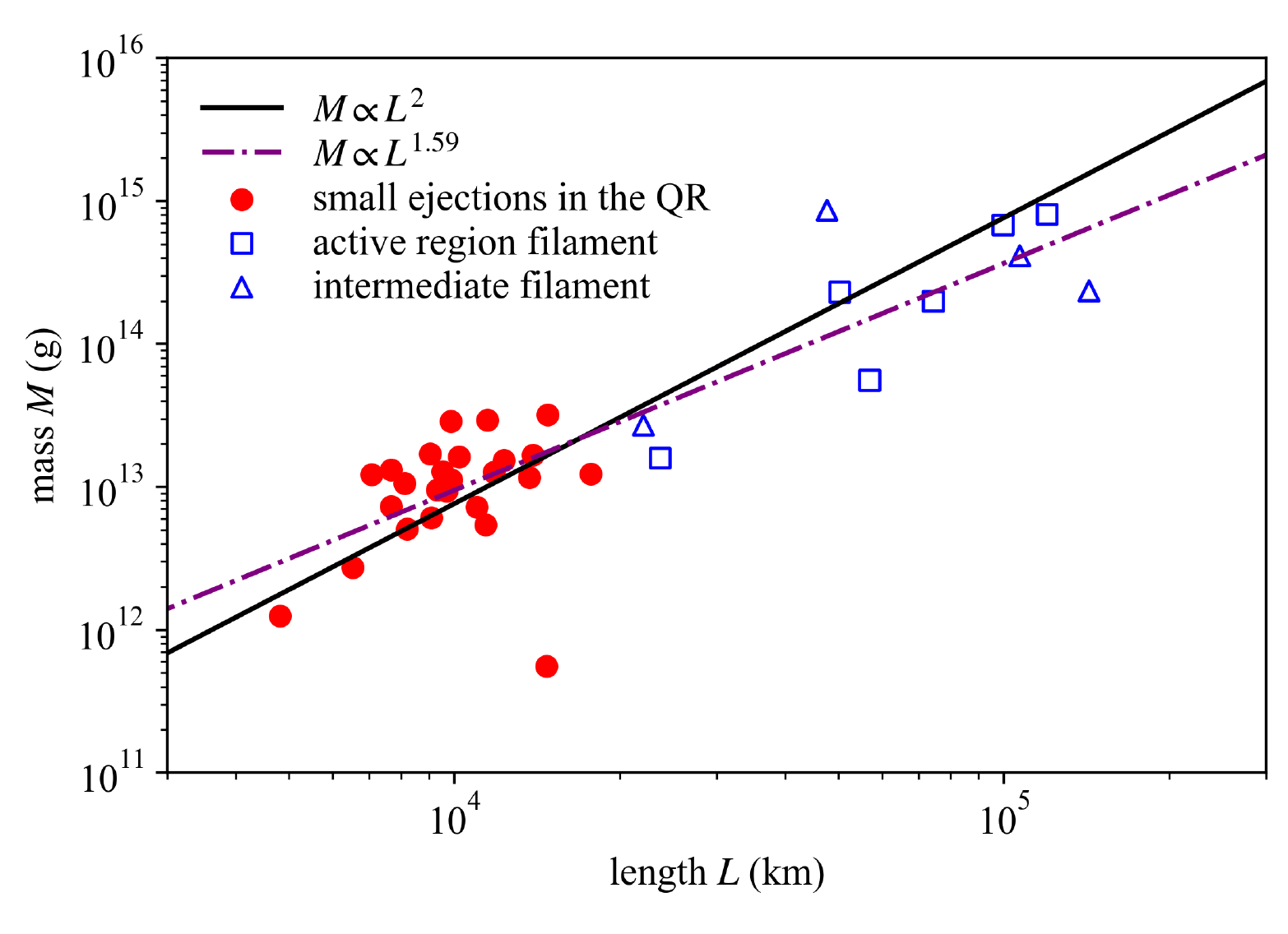}
\caption{
The relation between filament length and mass.
The red circles, blue squares, and blue triangles represent small mass ejections in the quiet region, the active region filament, and the intermediate filament eruptions, respectively.
The solid black line show the fitting result assuming $M\propto L^{2}$.
The dash-dotted purple line shows the fitting of all events in Figure \ref{fig:M_L} without fixing the power-law index .
}\label{fig:M_L}
\end{figure}

To explain the relationship between the mass of the cold plasma ejections and the total flare energy, we attempt to derive a scaling law using a simple model.
We assume that the filament is approximated by a cuboid supported by a stable helical magnetic field shown in Figures \ref{fig:model}a and \ref{fig:model}b.
In this situation, the characteristic spatial scale of the filament width $l$ is proportional to the scale height due to gravitational stratification along the concave magnetic field.
On the other hand, filament length $L$ and height $R$ are considered to vary similarly with the spatial scale. 
Hence, we can infer $M=\alpha L^2$ ($\alpha =$ const in units of $\,\mathrm{g}\,\mathrm{cm}^{-2}$). 
Since we determined both mass and length of the cold ejecta from our observation (Figures \ref{fig:width_estimate} and \ref{fig:all}), we can determine the coefficient $\alpha$ by fitting of the data analyzed in this study.
We show the relation between filament length and mass in Figure \ref{fig:M_L}, and the fitting result is $\alpha=10^{-5.12\pm0.079}$.
Note that if the fitting is performed without assuming the $M\propto L^{2}$ relation, we obtain $M\propto L^{1.59\pm0.18}$.
We also assume that a stable filament erupts with a flare, as shown in Figure \ref{fig:model}c.
In this case, the total flare energy can be estimated by the total amount of magnetic energy contained in a cube of length $L$.
Hence, using the conversion rate $f$ from the magnetic energy, the total flare energy can be roughly expressed as follows:
\begin{equation}
E_{\mathrm{tot}}= f\frac{B_{\mathrm{corona}}^2}{8\pi}L^3, \label{E_tot}
\end{equation}
where $B_{\mathrm{corona}}$ is the strength of the coronal magnetic field surrounding the filament.
Using $M=10^{-5.12} L^2$, we can derive the scaling law as follows;
\begin{eqnarray}
M=8.8\times 10^{12}\,  \left(\frac{f}{0.1}\right)^{-2/3} \left(\frac{B_{\mathrm{corona}}}{50\,\mathrm{G}}\right)^{-4/3}\left(\frac{E_{\mathrm{tot}}}{10^{28}\,\mathrm{erg}}\right)^{2/3}.\label{MvsE_2}
\end{eqnarray}

We show the scaling law (\ref{MvsE_2}) for the cases $B_{\mathrm{corona}}= 5 \,\mathrm{G}$ and $B_{\mathrm{corona}}= 50 \,\mathrm{G}$ in Figure \ref{fig:all}b.
We confirm from Figure \ref{fig:all}b that the scaling law explains well the present observation and also stellar events.
This result implies that cold plasma ejections with flares occur by a common mechanism regardless of their scale, and supports the interpretation of blue shifts associated with stellar flares as stellar filament eruptions \citep{2021NatAs...6..241N}.
We can also understand from the scaling law that the power on the total energy becomes smaller if we include large flares, as shown in Figure \ref{fig:all}a; the reason could be the difference of the coronal magnetic field strength in quiet and active regions.

We point out the similarity between the scaling law of cold plasma ejection and that of CMEs.
\citet{2016ApJ...833L...8T} assume a cube of spatial scale $L_{\mathrm{corona}}$ corresponding to the characteristic length of the active region in the gravitationally stratified corona, and its mass is supposed to be the CME mass ($M_{\mathrm{CME}}$).
Therefore, the CME mass can be estimated as $M_{\mathrm{CME}}\sim\rho_{\mathrm{corona}} L_{\mathrm{corona}}^2H_{\mathrm{corona}}\propto L_{\mathrm{corona}}^2$, where $\rho_{\mathrm{corona}}$ is the mass density at the bottom of the corona and $H_{\mathrm{corona}}$ is the scale height in the corona.
As a result, the mass is proportional to the 2/3 power of the total flare energy, similar to that in the cold plasma ejection case.
Although the shape of the ejected plasma is different in these theories, the scaling laws of cold plasma ejections and CMEs both follow $2/3$ power.
This is because we can assume that, even for the hot plasma, the density scale height is smaller than the typical length of coronal magnetic field lines.

Next, we attempt to derive a scaling law by estimating the filament width and height to understand what physical quantity affects the coefficient $\alpha$ of the relation $M\propto L^2$.
Based on the calculations in appendix \ref{sec:theory}, we derive the following scaling law:
\begin{eqnarray}
M=&&9.0\times 10^{12}\, \left(\frac{f}{0.1}\right)^{-2/3} \left(\frac{B_{\psi}/B_y}{0.3}\right) \left(\frac{\beta_x(x=0)}{10^{-3}}\right)^{-1/2}\nonumber \\ 
&&\left(\frac{H}{250\,\mathrm{km}}\right) \left(\frac{\rho}{3\times10^{-14}\,\mathrm{g}\,\mathrm{cm}^{-3}}\right)
\left(\frac{B_{\mathrm{corona}}}{50\,\mathrm{G}}\right)^{-4/3}\left(\frac{E_{\mathrm{tot}}}{10^{28}\,\mathrm{erg}}\right)^{2/3}, \label{MvsE}
\end{eqnarray}
where $B_{\psi}$ is an azimuthal component of the helical magnetic field and $\beta_x(x=0)$ is the plasma $\beta$ (the ratio of gas pressure to magnetic pressure) calculated from only the horizontal magnetic field $B_x$ in the filament center.
$H$ and $\rho$ are the filament's scale height and mass density.
In deriving the above scaling law, we can express the coefficient $\alpha$ as in equation (\ref{alpha_app}).
Substituting the same values as in equation (\ref{MvsE}) into equation (\ref{alpha_app}), the coefficient $\alpha$ can be expressed as follows:
\begin{eqnarray}
\alpha = 9.1\times10^{-6}\left(\frac{B_{\psi}/B_y}{0.3}\right)\left(\frac{\beta_x(x=0)}{10^{-3}}\right)^{-1/2}\left(\frac{H}{250\,\mathrm{km}}\right)\left(\frac{\rho}{3\times10^{-14}\,\mathrm{g}\,\mathrm{cm}^{-3}}\right) . \label{alpha_th}
\end{eqnarray}
The value of $\alpha$ determined by the parameter given in equation (\ref{alpha_th}) is consistent with the value determined by fitting from the observed data $(=10^{-5.12})$.

For M-type stars, some parameter values of the scaling law, such as $B_{\mathrm{corona}}$, $H$, and $\rho$, should be different from those of denominators in equation (\ref{MvsE}) because the dipolar magnetic field is expected to be 10--1000 times stronger than that of Sun-like stars \citep{2017MNRAS.466.1542S} and the gravitational acceleration $g$ and surface temperature are also different on these stars.
These differences may affect roughly one order of magnitude on the scaling law and may explain the much smaller ejected mass in some events.
Regarding the magnetic field strength, we need the coronal magnetic field strength, not the photospheric magnetic field strength, to use our scaling law.
Thus, the value of the magnetic field in M-type stars used in our scaling law should be several times smaller than the dipole field values  (typically 200 G, up to 2000 G) summarized in \citet{2017MNRAS.466.1542S}.
Assuming that the coronal magnetic field is three times smaller than the photospheric field, we obtain about 60-600 G as the coronal magnetic field of an M-type star.
These coronal field values can explain why the ejected mass of some events appear as extensions of the solar case while some are about an order of magnitude smaller in Figure \ref{fig:all}b.
This estimate is also roughly consistent with the M-type star's coronal magnetic field strength estimated by the scaling law between flare duration and flare energy  \citep{2017ApJ...851...91N,2021PASJ...73...44M} and  the scaling law between flare temperature and flare emission measure \citep{1999ApJ...526L..49S,2002ApJ...577..422S,2007MNRAS.379.1075R}.
Meanwhile, according to \citet{2021ApJ...919...29S}, $\rho H=p/g$ in the transition region changes only a factor of three approximately between the Sun and M-type stars.
The calculation of \citet{2021ApJ...919...29S} was for an open flux tube with a different geometry from a closed loop.
However, assuming the same trend in the closed loop and that the gas pressure in the filament and the transition region is comparable, the difference in $\rho H$ on the scaling law is expected to be a factor of three approximately.
From these discussions, we expect that equation (\ref{MvsE}) (and equation (\ref{MvsE_2})) are also applicable to M-type stars by accounting for differences in the coronal magnetic field strength.

\subsection{Kinetic Energy, Velocity, and Total Flare Energy Relation} \label{sec:sub_kinevel}

Here, we discuss the theoretical relationship among kinetic energy, velocity, and total flare energy spanning more than ten orders of magnitude,  as shown in Figures \ref{fig:all_kine} and  \ref{fig:all_vel}.
For this purpose, we consider the relations for solar CMEs ($E_{\mathrm{kin}}\propto E_{\mathrm{tot}}$ and $v\propto E_{\mathrm{tot}}^{1/6}$), which \citet{2016ApJ...833L...8T} reported being established for solar CMEs, and compare these relations with the present analysis for cold plasma ejections.

The relationship $v\propto E_{\mathrm{tot}}^{1/6}$ is roughly consistent with the upper limits of the velocity for each energy in Figure \ref{fig:all_vel}.
Note that we used the “line-of-sight" velocity in this observation, which provides a lower bound of the velocity due to the projection effect.
Hence, the theoretical $v\propto E_{\mathrm{tot}}^{1/6}$ relationship can appear only for the upper limits for each total flare energy in Figure \ref{fig:all_vel}.
The relationship $v\propto E_{\mathrm{tot}}^{1/6}$ may correspond to a larger Alfv\'{e}n  velocity as the spatial scale increases.
In other words, it would indicate that as the spatial scale increases, the height of the ejecta increases, and the plasma density decreases.

Figure \ref{fig:all_kine} shows that $E_{\mathrm{kin}}=0.1 E_{\mathrm{tot}}$ is a good approximation for the relation between the total flare energies and the kinetic energy of the cold plasma ejections in a range of $E_{\mathrm{tot}}\sim10^{27}-10^{29}\,\mathrm{erg}$.
However, the power obtained by our observation ($E_{\mathrm{kin}}\propto E_{\mathrm{tot}}^{0.68}$ in Figure \ref{fig:all_kine}) is less than one for a wider range of energies.
 For small events in the quiet region, the ratio of kinetic energy to total energy is greater than 10\%, and for some stellar cold plasma ejections, the ratio is significantly smaller than 10\%.
Therefore, it is inconsistent with our observation for a wide energy range, although the $E_{\mathrm{kin}}=0.1 E_{\mathrm{tot}}$ relationship is consistent with the relationships $M\propto E_{\mathrm{tot}}^{2/3}$ and $v\propto E_{\mathrm{tot}}^{1/6}$.

Are there any problems with our analysis method that prevented $E_{\mathrm{kin}}=0.1 E_{\mathrm{tot}}$ from being valid for a wide range of energies?
For small events in the quiet region, neglecting non-thermal energy may cause the underestimation of the total flare energy.
In addition, recent studies have suggested that only a portion of the loops are brightened in EUV seen in coronal temperature in small flares in the quiet region \citep{2021A&A...656L...4B,2021A&A...656L...7C}.
In other words, even if magnetic reconnection occurs, the major part of the loop may remain at the chromospheric temperature.
Since the thermal energy is obtained from the EUV in our observation, we cannot measure the thermal energy of the plasma at chromospheric temperature, which may lead to an underestimation of the total flare energy.

The smaller kinetic energies obtained for some stellar events than that expected from the $E_{\mathrm{kin}}=0.1 E_{\mathrm{tot}}$ relation may reflect the unique nature of stellar ejections.
Some numerical studies for stellar CMEs have reported the possibility of the stellar CMEs suppression by a large-scale stellar coronal magnetic field \citep{2018ApJ...862...93A,2022MNRAS.509.5075S}.
In addition, the spatial scale of the stellar ejection is comparable to the stellar radius, and its spectra may be more sensitive to the projection effect than in solar cases. 
Consequently, the line-of-sight velocity and the kinetic energy are more likely to be underestimated in the stellar cases than in the solar cases. 
To understand the significance of this effect, we need further sun-as-a-star analyses for the spectra of the cold plasma ejection \citep{2021NatAs...6..241N,2022arXiv221002819O}.

\subsection{Error Sources and Validity of Event Selection Methods} \label{sec:sub_kinevel}

We note that our estimation of the physical quantities has errors of approximately one order of magnitude. 
When estimating ejection mass, there are uncertainties in the line-of-sight thickness of ejecta and the validity of selecting the time and pixels at which the mass was estimated.
We have analyzed the same event as \citet{2021NatAs...6..241N} and have confirmed that our mass estimation is about factor 2 smaller than their result due to the selection of the time and pixels (Figure \ref{fig:all}b).  
Uncertainties also arise from the model used to estimate the mass \citep{1997A&A...324.1183T}.
Thus, the ejection mass is expected to have an error of approximately one order of magnitude.
In estimating the energy of small flares in the quiet region, we used their thermal energy based on \citet{2020A&A...644A.172W}.
Flare energies accompanied by filament eruption were estimated by assuming that the bolometric energy is proportional to the GOES flare class. 
However, the flare energy partition has not been concretely established, and some studies have suggested that the power between bolometric energy and GOES flare class has different values \citep[see][Fig. 22]{2022LRSP...19....2C}.
In addition, the estimated kinetic energy is supposed to be a lower bound. 
This underestimation is due to our measurement of only line-of-sight velocity and neglecting the kinetic energy of the hot coronal plasma.
For determining the flare energy in the M-type stars, the bolometric energy was estimated from the X-ray energy, which was estimated from the H$\alpha$ line energy.
Therefore, the energy uncertainty is considered more significant than the solar cases.


We note that transverse motions of plasma during the wavelength scan will not affect the results of our analysis.
SDDI can take 73 images from H$\alpha-9.0$ \AA\, to H$\alpha+9.0$ \AA\, in ten seconds  \citep{2017SoPh..292...63I}.
Hence, the time difference in observing from H$\alpha-2.0$ \AA\, to H$\alpha+2.0$ \AA\, is approximately $2\,\mathrm{s}$.
Since the typical line-of-sight velocity of the plasma analyzed in this study is $30\,\mathrm{km}\,\mathrm{s^{-1}}$ (Figure \ref{fig:all_vel}), the plasma moves horizontally by approximately $(2\,\mathrm{s}\times 30\,\mathrm{km}\,\mathrm{s^{-1}})/(1.23\,\mathrm{arcsec} = 8.99\times10^2\,\mathrm{km})=1/15$ of the SDDI pixel size during a single scan in the wavelength direction. Moreover, we performed spatial averaging in deriving the physical quantities in this study. Based on the above discussions, the effect of observing different plasmas with the same pixel would be minor in this study.

Our analysis results are expected to contain errors of approximately one order of magnitude; however, the relationships between physical quantities over a wide range of total flare energies do not vary significantly.
In the fitting results by the linear regression method in this study, the errors in power index were estimated to be about 0.1 in all cases ($M\propto E_{\mathrm{tot}}^{0.46\pm0.06}$,  $E_{\mathrm{kin}}\propto E_{\mathrm{tot}}^{0.68\pm0.12}$, and $v\propto E_{\mathrm{tot}}^{0.10\pm0.04}$).
Consequently, the physical quantity predicted from our fitting results are approximately one order of uncertainty for ten orders of magnitude variation in total flare energy.
On the other hand, our linear regression method did not account for errors in the values of the $x$-axis, that is, the total flare energy.
We also estimated the power indexes and their errors by the orthogonal distance regression (ODR) method to consider the error in both and $x$- and $y$-axes.
As a result of the fitting with an error of one order of magnitude in the total flare energy, we obtained $M\propto E_{\mathrm{tot}}^{0.54\pm0.05}$, $E_{\mathrm{kin}}\propto E_{\mathrm{tot}}^{0.75\pm0.07}$, and $v\propto E_{\mathrm{tot}}^{0.11\pm0.02}$.
These results were almost similar to those obtained using linear regression methods.
We also fitted using the ODR method for the relationship between the cold plasma's length and mass, assuming a factor two error in the length.
We obtained $M=10^{-5.12\pm0.07} L^{2}$ and $M\propto L^{1.9\pm0.18}$, and these results were similar to those by  linear regression methods.
Note that we assumed symmetric errors for the fittings in this study.
However, some of our physical quantities estimation errors only act in one direction: asymmetric errors (such as underestimation of kinetic energy and neglection of non-thermal energy for small flares in the quiet region).
Therefore, while our analysis leaves open the issue of handling asymmetric errors, it is believed to reflect the correct trend as long as the errors are symmetric.

Our analysis excluded events that have only cold plasma ejections or only flare brightening.
Because of this, we can perform comparisons with the stellar cases.
If we include events in which only cold plasma ejections occur, it is formally possible to add them in Figures \ref{fig:all} through \ref{fig:all_vel} by setting the flare thermal energy or bolometric energy to the detection limit in our method.
Because kinetic energy accounts for a large proportion of the total flare energy in small events in the quiet region, those events will be similar to our analysis results.
In contrast, for large filament eruptions, the flare energy would shift toward an order of magnitude or two smaller.
Events with only flares cannot be added to Figures \ref{fig:all} through \ref{fig:all_vel} because the physical quantities of the cold plasma ejecta are not present.
Some flares are accompanied by hot plasma eruptions at coronal temperature (plasmoid ejections) seen in soft X-ray or EUV without cold plasma ejections \citep[e.g.,][]{1998ApJ...499..934O,2012ApJ...745L...6T}.
In this sense, the correlations and scaling laws for mass and kinetic energy of cold ejecta investigated in this study would correspond to an upper limit on the low-temperature component in eruptive events.
\citet{1998ApJ...499..934O} reported the mass of the hot plasmoid and its kinetic energy (mass: $(2.3\pm0.2)\times10^{13}\,\mathrm{g}$, kinetic energy: $8\times10^{27}-2\times10^{28}\,\mathrm{erg}$) close to those obtained in this study for an M2.0 class flare.
Those hot ejections are  desired to be included to establish a complete scenario on the relationship between flares and plasma ejections.

\vskip\baselineskip
In this study, we estimated  physical parameters of various scale solar cold plasma ejection accompanied by flares.
We showed their positive correlations and constructed a theoretical scaling law between the total flare energy and the ejected mass.
Our scaling law suggests a common physical mechanism across a wide range of energies and can also be used to estimate parameters such as magnetic fields indirectly.
In particular, we expect to make more progress in parameter estimation for stellar flare observations when combined with other scaling laws \citep{1999ApJ...526L..49S,2002ApJ...577..422S,2017ApJ...851...91N}.
We expect that our results will serve as a quantitative benchmark for studies of stellar flares \citep{2021NatAs...6..241N} and “campfires" associated with cold plasma ejection \citep{2021ApJ...921L..20P}.

\begin{acknowledgments}
We wish to thank the anonymous referee for helpful comments that led to improvements in this work.
We thank T. Yokoyama, Y. Notsu, and K. Namekata for fruitful discussions and comments.
We are grateful to the staff of Hida Observatory and K. Hirose for the instrument development, daily observations, and finding many eruptive events in the quiet region.
We are grateful to the SDO/AIA teams. 
SDO is part of NASA’s Living with a Star Program. 
We want to thank Editage (www.editage.com) for English language editing.
This research is supported by JSPS KAKENHI grant numbers 22J14637 (Y.K.), 21H01131 (K.S., K.I., and A.A.), and 21J14036 (D.Y.).
\end{acknowledgments}

\software{
sunpy \citep{2020ApJ...890...68S},
aiapy\citep{2020JOSS....5.2801B},
NumPy\citep{2020NumPy-Array},
SciPy\citep{2020SciPy-NMeth}
 }

\appendix

\section{Derivation of the Scaling Law for Filament Mass and Total Flare Energy} \label{sec:theory}

In this section, we attempt to establish a theoretical relationship between the filament’s mass and the total flare energy. 
For this purpose, we evaluate the approximate values of the filament mass and total flare energy in the form of the filament length $L$.
We begin to discuss the formula for evaluating the stably existing filament mass.

We assume that the filament is approximated by a rectangle of length $L$, height $R$, and width $l$ (Figure \ref{fig:model}).
A stable helical magnetic field supports the filament.
The radius of the helical field gives the height $R$ of the filament.

For this helical spiral to be stable against the kink instability, $R$ should be greater than that determined by the following Kruskal--Shafranov limit \citep{1954RSPSA.223..348K,2014masu.book.....P}.
\begin{equation}
R > \frac{L^{'}}{2\pi }\frac{B_{\psi}}{B_y}, \label{KS_con}
\end{equation}
where $L^{'}$ is the length required for the helical magnetic field lines to make one rotation.
From inequality (\ref{KS_con}), we assume that $R$ is proportional to $L$ for the filament to exist stably.
$L^{'}$ should be proportional to the filament length $L$.
Thus, we assume that the radius $R$ of a stably existing helical field can be expressed using the filament length $L$ as follows:
\begin{equation}
R=a\frac{L}{2\pi }\frac{B_{\psi}}{B_y} =\frac{a}{4}\frac{L}{\pi/2 }\frac{B_{\psi}}{B_y} , \label{KS_con_R}
\end{equation}
where $a>1$, $B_y$, and $B_{\psi}$ are a constant, magnetic field strength in the filament axial direction, and an azimuthal component of the magnetic field, respectively.
If we assume that $B_{\psi}/B_y=0.3$ is constant in the helical field and $L\sim 5\times 10^4 \,\mathrm{km}$, we obtain $R\sim 9.6\times 10^3 \times a/4\,\mathrm{(km)}$.
Thus, we set $a$ to four to be consistent with the typical prominence height.

To estimate the filament width  $l$, we use the analytical solution of  the Kippenhahn--Schl\"{u}ter model \citep{1957ZA.....43...36K} only near the bottom of the helical magnetic field. 
Kippenhahn--Schl\"{u}ter model assumes that the filament is in hydrostatic equilibrium and $B_x$, $B_y$, and temperature $T$ are constant in the $x$ direction.
In this model, the filament gas pressure $p$ and the vertical component of magnetic field $B_z$ are expressed as follows \citep{2014masu.book.....P}:
\begin{eqnarray}
&p& =\frac{B_{z\infty}^2 - B_z^2}{8\pi} \label{Kip-Sch_p} \label{Kip-Sch_p}\\
&B_z& =B_{z\infty}\tanh{\frac{B_{z\infty}x}{2B_x H} }, \label{Kip-Sch_Bz}
\end{eqnarray}
where $H$ is the scale height of the filament and $B_{z\infty}=B_z(x=+\infty)$.
We assume that the filament width $l$ can be approximated by the typical spatial scale of gas pressure distribution (\ref{Kip-Sch_p}).
\begin{equation}
 l=\frac{2B_x H}{B_{z\infty}}. \label{KS_width}
 \end{equation}
Based on $B_{z}(x=0)=0$, $B_{x}/B_{z\infty}$ can be represented using equation (\ref{Kip-Sch_p}) as follows:
\begin{equation}
\frac{B_x}{B_{z\infty}} = \left(\frac{B_x^2}{8\pi p(x=0)}\right)^{1/2} =\beta_x^{-1/2}(x=0) , \label{Kip-Sch_betax}
\end{equation}
where $\beta_x^{-1/2}(x=0)$ is the plasma $\beta$ calculated from only $B_x$ in the filament center.
Thus, the filament width can be written as follows:
\begin{equation}
l =2\beta_x^{-1/2}(x=0)H. \label{Kip-Sch_l} 
\end{equation}

From the above discussion, using the filament typical density $\rho$, the filament mass $M$ can be expressed as follows:
\begin{eqnarray}
&M&= LRl\rho =  \frac{4\beta_x^{-1/2}(x=0)H }{\pi}\frac{B_{\psi}}{B_y}\rho L^2=\alpha L^2,  \label{FM} \\
&\alpha& = \frac{4\beta_x^{-1/2}(x=0)H }{\pi}\frac{B_{\psi}}{B_y}\rho, \label{alpha_app}
\end{eqnarray}
where $\alpha $ is the coefficient when rewritten in the form $M\propto L^2$ and corresponds to the coefficient of the solid black line in Figure \ref{fig:M_L}.

By eliminating $L$ from equations (\ref{FM}) and (\ref{E_tot}) and assuming the parameter values, the following relationship between filament mass $M\,\mathrm{(g)}$ and total flare energy $E_{\mathrm{tot}}\,\mathrm{(erg)}$ can be obtained:
\begin{eqnarray}
M=&&9.0\times 10^{12}\, \left(\frac{f}{0.1}\right)^{-2/3} \left(\frac{B_{\psi}/B_y}{0.3}\right) \left(\frac{\beta_x(x=0)}{10^{-3}}\right)^{-1/2}\nonumber \\ 
&&\left(\frac{H}{250\,\mathrm{km}}\right) \left(\frac{\rho}{3\times10^{-14}\,\mathrm{g}\,\mathrm{cm}^{-3}}\right)
\left(\frac{B_{\mathrm{corona}}}{50\,\mathrm{G}}\right)^{-4/3}\left(\frac{E_{\mathrm{tot}}}{10^{28}\,\mathrm{erg}}\right)^{2/3}. \label{MvsE_ap}
\end{eqnarray}
We determined the parameters $f$ of the above equation (\ref{MvsE_ap}) from the ratio between the magnetic energy related to the flare and the bolometric energy  \citep{2012ApJ...759...71E}.
In addition, for the above parameters, $\alpha$ in equation (\ref{alpha_app}) takes the following values:
\begin{eqnarray}
\alpha = 9.1\times10^{-6}\left(\frac{B_{\psi}/B_y}{0.3}\right)\left(\frac{\beta_x(x=0)}{10^{-3}}\right)^{-1/2}\left(\frac{H}{250\,\mathrm{km}}\right)\left(\frac{\rho}{3\times10^{-14}\,\mathrm{g}\,\mathrm{cm}^{-3}}\right) . \label{alpha_th_app}
\end{eqnarray}

The range of parameters  $B_{\psi}/B_y$, $\beta_x(x=0)$, $H$, and $\rho$ is estimated to be approximately the following degrees, from which we select appropriate values.
As for the orientation between the filament axis and the magnetic field orientation, \citet{2017ApJ...851..130H} said that it is concentrated between $10^{\circ}$ and $30^{\circ}$ in both active and quiet regions.
From this relationship, we can infer that $B_{\psi}/B_y\sim \tan10^{\circ}$--$30^{\circ}\sim0.176$--$0.577$.
Previous studies have reported prominence electron temperatures $T\,\mathrm{(K)}$, electron density $n_e\,\mathrm{(cm^{-3})}$, hydrogen ionization ratio $\chi$, and gas pressure $p\,(\mathrm{dyn}\,\mathrm{cm^{-2})}$ of $4000<T<10000$, $10^9<n_e<10^{11}$, $0.2<\chi<0.9$, and $0.02<p<1$ \citep{2010SSRv..151..243L}.
Assuming mean molecular weight $1/\mu=1.5$, we can estimate scale height $H=R_gT/\mu g =180$ -- $450\,\mathrm{(km)}$, where $R_g=8.31\times 10^7\,\mathrm{(erg}\,\mathrm{mol}^{-1}\,\mathrm{K}^{-1})$ is the gas constant and $g=2.74\times10^4\,\mathrm{(cm}\,\mathrm{s}^{-1})$ is the solar gravitational acceleration.
In addition, $\rho=\mu m_{\mathrm{H}}n \simeq \mu m_{\mathrm{H}}(n_{\mathrm{H}} + n_{\mathrm{He}} + n_e)=\mu m_{\mathrm{H}}(n_{\mathrm{H}}/n_e + n_{\mathrm{He}}/n_{\mathrm{H}}\times n_{\mathrm{H}}/n_e + 1)n_e$.
Assuming $n_{\mathrm{He}}/n_{\mathrm{H}} = 0.08$ and $n_{\mathrm{H}}/n_e\simeq\chi$, we can derive $\rho\simeq 1.11\times10^{-24}(1.08\chi  + 1)n_e$.
The above equation shows that the ionization ratio affects only about factor 2, so assuming $\chi = 0.5$, we can estimate $\rho \simeq1.7\times10^{-15}$ -- $1.7\times10^{-13}\,(\mathrm{g}\,\mathrm{cm}^{-3})$.
The magnetic field strength of the quiescent prominence is estimated to be about 10 -- 80 G \citep{2003ApJ...598L..67C,2005ApJ...622.1265C}.
Assuming that the horizontal component is $B_x(x=0)=5$ -- $40\,\mathrm{G}$, we can estimate $\beta_x(x=0)=0.02/(40^2/8\pi)$ -- $1.0/(5^2/8\pi)\simeq 3\times10^{-4}$--$1.0$ in the quiet region.
On the other hand, the magnetic field strength of the active region prominence is estimated to be around 100 -- 800 G \citep{2009A&A...501.1113K,2011A&A...526A..42S,2012ApJ...749..138X}.
Thus, assuming that the horizontal component is $B_x(x=0)=50$ -- $400\,\mathrm{G}$, we can estimate $\beta_x(x=0)=0.02/(400^2/8\pi)$ -- $1.0/(50^2/8\pi)\simeq 3\times10^{-6}$--$1.0\times10^{-2}$ in the active region.
A more precise determination of these parameters in future filament observations would increase the reliability of our scaling law.


\bibliography{draft_bb}
\bibliographystyle{aasjournal}



\end{document}